\documentclass[%
 reprint, nobibnotes,
 amsmath,amssymb,prx,
 aps]{revtex4-2}

\usepackage{graphicx}% Include figure files
\usepackage{dcolumn}% Align table columns on decimal point
\usepackage{bm}% bold math
\usepackage{multirow}
\usepackage{makecell}

\begin{document}

\title{Monolithically integrated, broadband, high-efficiency silicon nitride-on-silicon waveguide photodetectors in a visible-light  integrated photonics platform}% Force line breaks with \\

\author{Yiding Lin\textsuperscript{1}} \altaffiliation[Currently at ]{Institute of Microelectronics, Agency for Science, Technology and Research (A*STAR), 138634, Singapore.} \email{yidinlin@mpi-halle.mpg.de}
\author{Zheng Yong\textsuperscript{2}}%
\author{Xianshu Luo\textsuperscript{3}}
\author{Saeed Sharif Azadeh\textsuperscript{1}}
\author{Jared Mikkelsen\textsuperscript{1}}
\author{Ankita Sharma \textsuperscript{1,2}}
\author{Hong Chen\textsuperscript{1}}
\author{Jason C. C. Mak\textsuperscript{2}}
\author{Patrick Guo Qiang Lo\textsuperscript{3}}
\author{Wesley D. Sacher\textsuperscript{1}}
\author{Joyce K. S. Poon\textsuperscript{1,2}}
 \email{joyce.poon@mpi-halle.mpg.de}

 \affiliation{\textsuperscript{1}Max Planck Institute of Microstructure Physics, Weinberg 2, 06120 Halle, Germany}%
\affiliation{\textsuperscript{2}Department of Electrical and Computer Engineering, University of Toronto, 10 King’s College Road, Toronto, Ontario M5S 3G4, Canada}
\affiliation{\textsuperscript{3}Advanced Micro Foundry Pte Ltd, 11 Science Park Road, Singapore Science Park II, 117685, Singapore}

\date{\today}% It is always \today, today,

\begin{abstract}
Visible and near-infrared spectrum photonic integrated circuits are quickly becoming a key technology to address the scaling challenges in quantum information and biosensing. Thus far, integrated photonic platforms in this spectral range have lacked integrated photodetectors. Here, we report the first silicon nitride-on-silicon waveguide photodetectors that are monolithically integrated in a  visible light  photonic platform on silicon. Owing to a leaky-wave silicon nitride-on-silicon  design, the devices achieved a high external quantum efficiency of $>60\%$ across a record wavelength span from $\lambda \sim 400$ nm to $ \sim 640$ nm, an opto-electronic bandwidth up to 9 GHz, and an avalanche gain-bandwidth product up to $173 \pm 30$ GHz. As an example, a photodetector was integrated with a wavelength-tunable microring in a single chip for on-chip power monitoring.  
\end{abstract}

\maketitle

Silicon (Si) photonics leverages microelectronics fabrication in foundries to mass manufacture dense and complex photonic integrated circuits (PICs). Extending Si photonics into the visible and near-infrared (NIR) spectrum ($\lambda \sim 400-800$ nm) \cite{PorcelOLT2019, SacherOE2019, SoraceJSTQE2019, WestAPLPh2019, DongNP2022}  can bring PICs to an even wider range of emerging applications, including spectroscopy and flow cytometry \cite{GeuzebroekSAB2016, SubramanianPR2015}, neurophotonics \cite{SacherNeurophotonics2021, MoreauxNeuron2020, MohantyNatBE2020}, quantum information processing \cite{DongNP2022, MehtaNAT2020,NiffeneggerNAT2020}, underwater communication \cite{ShenOE2016} and scanning displays \cite{RavalOL2018, NotarosCLEO2019, ShinOL2020}. In Si photonics for the visible and NIR wavelength range, the optical waveguide is typically formed in a silicon nitride (SiN) or aluminum oxide (Al$_2$O$_3$) layer surrounded by a silicon dioxide (SiO$_2$) cladding on a Si or silicon-on-insulator (SOI) substate. Thus far, on 100, 200, or 300-mm diameter Si wafers, visible/NIR photonic platforms with SiN \cite{LioniX, MashayekhOE2021, HoffmanPJ2016, IMEC_BioPIX, RomeroGarciaOE2013, SoraceJSTQE2019, DongNP2022} or Al$_2$O$_3$ \cite{SoraceJSTQE2019, WestAPLPh2019} waveguides are usually passive, containing components such as gratings, power splitters, and interferometers, with  waveguide phase-shifters tuned by the thermo-optic or strain-optic effect \cite{DongNP2022, LiangNP2021, YongOE2022, DongARXIV2022}. Post-fabrication processing and heterogeneous integration steps are used to incorporate liquid crystal phase-shifters \cite{NotarosOE2022}, lasers  \cite{Corato-ZanarellaCLEO2021, SiddarthARXIV2021}, and photodetectors (PDs) \cite{GherabliARXIV2021,CuyversOL2022} with SiN waveguides.

On-chip optical-to-electrical conversion is an essential functionality in a PIC platform. In visible and NIR Si photonics, the Si substrate serves as a natural candidate for light absorption compatible with monolithic integration, circumventing heterogeneous integration that increases the fabrication complexity.  Si PDs are conventionally surface incident, but in PICs, waveguide PDs are preferred since they can be easily integrated with other in-plane circuit components.  The main challenge in realizing a broadband, high-efficiency waveguide PD is the input waveguide-to-Si coupling, which can exhibit significant wavelength dependence across the visible and NIR spectral bands. One approach is to use the SOI device layer for photodetection and an overhead waveguide to achieve leaky-wave coupling. However, such designs are complicated by the presence of multiple Si slab modes that hybridize with the waveguide mode when their effective indices are similar, resulting in resonance-like peaks in the absorption spectrum at wavelengths where the phase-matching condition is met \cite{AhnJAP2011,Morgan2021,ChatterjeeOL2019}. Another approach is to end-fire couple the input waveguide to a Si absorption region as in \cite{YanikgonulNATCOMM2021}, but this complicates the fabrication by requiring the input waveguide and Si to be co-planar, and the insertion loss is wavelength-dependent due to the spectral variation of the mode field diameter mismatch between SiN and Si.

\begin{figure*}[t]
\includegraphics[width = \textwidth]{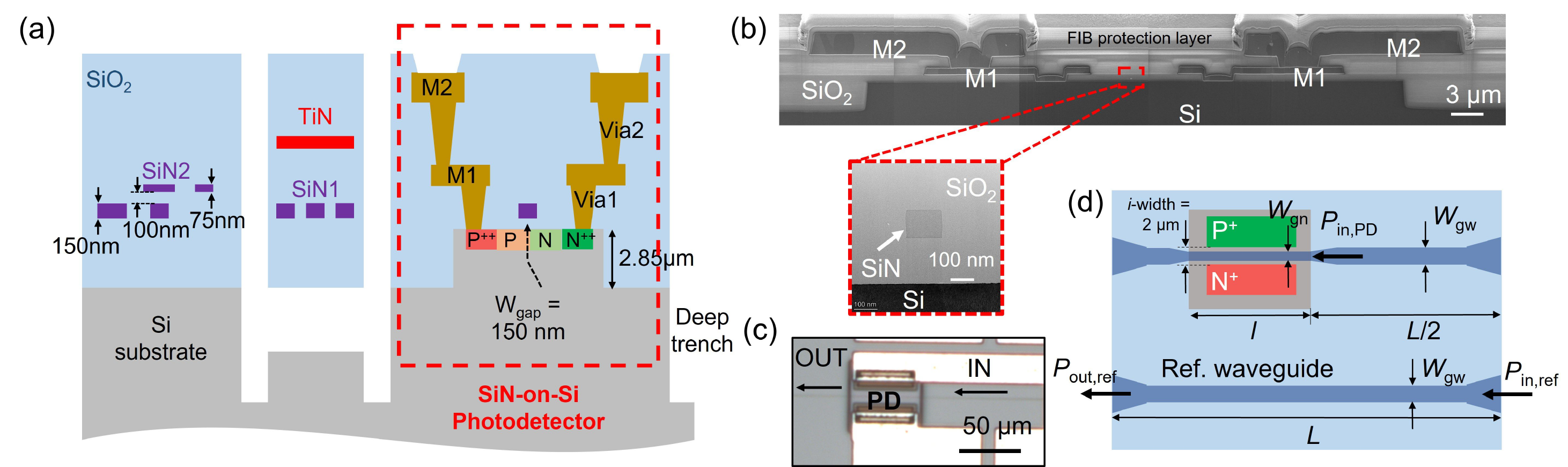}
\caption{SiN-on-Si waveguide PDs. (a) Cross-section of the visible spectrum integrated photonic platform on 200-mm Si. The platform consists of 2 SiN waveguide layers, a Si undercut, 2 metal layers, titanium nitride (TiN) high resistivity heaters, P and N doped regions on a Si mesa for photodetection, and a deep trench for waveguide facets. The cross-section of the PD, which has a PN (in (a)) or PIN junction (intrinsic region width = 2 {\textmu}m, a top view shown in (d)), is demarcated by the red dashed box. (b) A stitched scanning electron micrograph of the PD cross-section. A protective layer had been applied for the focused ion beam (FIB) milling to prepare the sample for imaging. Inset: Transmission electron micrograph showing the SiN waveguide above the Si. (c) An optical micrograph of a PD with an input SiN waveguide. (d) Top-view schematic of the PD test structure with a reference waveguide for power calibration. }\label{fig:fig1}
\end{figure*}

Here, we report the first broadband, high-efficiency, SiN-on-Si waveguide PDs integrated within a foundry-fabricated visible spectrum PIC platform fabricated on 200-mm Si wafers. Preliminary results of the devices were reported in {\cite{Lin:21}}. Figure \ref{fig:fig1}(a) shows the PIC platform cross-section, which comprises 2 layers of SiN waveguides, a titanium nitride (TiN) heater, and metal layers on bulk Si. Components from this platform, such as the suspended thermo-optic phase-shifters, bi-layer fiber-to-chip couplers, and electro-thermally actuated micro-electro-mechanical system (MEMS) cantilevers, are reported elsewhere \cite{LinOE2021, YongOE2022, AzadehCLEO2022, AzadehARXIV2022}.  The SiN-on-Si PDs have a leaky-wave design where the optical power in the SiN input waveguide evanescently leaks into a mesa defined in bulk Si under the waveguide. The thick and wide Si mesa effectively supports a continuum of radiation modes, eliminating the coupling sensitivity to the phase-matching condition in SOI-based devices and facilitating broadband and efficient power transfer. The PD geometry is alignment tolerant. We achieved a record operating wavelength span of $>230$ nm, with an external quantum efficiency (EQE) $>60\%$ from $\lambda \sim 400$ to $\sim 640$ nm for the transverse magnetic (TM) polarization. The PDs exhibited opto-electronic (OE) bandwidths up to 9 GHz and also operated as avalanche photodiodes (APDs) with a gain-bandwidth product up to $173 \pm 30$ GHz. Lastly, as a proof of concept, we demonstrate the integration of the PD with a tunable microring filter in a single-chip for on-chip power monitoring. %Although this work focused on the visible spectrum, the PD design concept can also be applied to  photodetection in the NIR.

\section{Results}

The dashed box in Figure \ref{fig:fig1}(a) shows the cross-section of the PD design, which is comprised of a SiN waveguide passing atop a lateral PIN or PN junction in a Si mesa. Unless otherwise stated, all indicated dimensions are nominal values. For  PIN junctions, the intrinsic region had a width of 2 $\mathrm{\mu m}$, centered below the SiN waveguide. The gap ($W_{gap}$) between the SiN  and Si layer was designed to be 150 nm. Two layers of metals and vias connected the device to bond pads. The vias and metals were placed sufficiently apart to ensure optical isolation from the SiN waveguide. The Si mesa height was 2.85 $\mathrm{\mu m}$ to achieve a low propagation loss for the routing SiN waveguides (SiN1 layer in Fig. \ref{fig:fig1}(a)). Fig. \ref{fig:fig1}(b) shows a scanning electron micrograph of the PD cross-section with the SiN waveguide atop the Si mesa in the inset, and Fig. \ref{fig:fig1}(c) shows an optical micrograph of a device.

Figure \ref{fig:fig1}(d) shows the top-view schematic of a PD device with a reference waveguide structure for optical power calibration during characterization. As designed, the routing SiN waveguides had a thickness of $t=150$ nm and widths ($W_{gw}$) of 500 nm (for the single PD devices) and 380 nm (for the microring integration). Tapered edge couplers, starting with a SiN width of 5.2 $\mathrm{\mu m}$ that adiabatically narrowed to $W_{gw}$ over a length of 300 $\mathrm{\mu m}$, were used for fiber-to-chip coupling. The routing waveguides adiabatically narrowed over a length of 100 $\mathrm{\mu m}$ to $W_{gn}$ (i.e., 150, 200 and 250 nm) at the Si mesa facet for light to penetrate into the Si. We have previously reported the propagation loss of the routing waveguides ($\sim 1.1-5.9$ dB/cm from $\lambda$ = 430-648 nm {\cite{SacherOE2019}}, and $\sim 1.8-7.1$ dB/cm from $\lambda$ = 405-640 nm {\cite{LinOE2021}}) and insertion loss of the edge couplers (7.5-11.3 dB/facet from $\lambda$ = 430-648 nm {\cite{SacherOE2019}}, and 5.9-8.6 dB/facet from $\lambda$ = 405-640 nm {\cite{LinOE2021}}). The corresponding measurements for this work are in the Supplement Information (Fig. S2).

\begin{figure*}
\centering
\includegraphics[width=\textwidth]{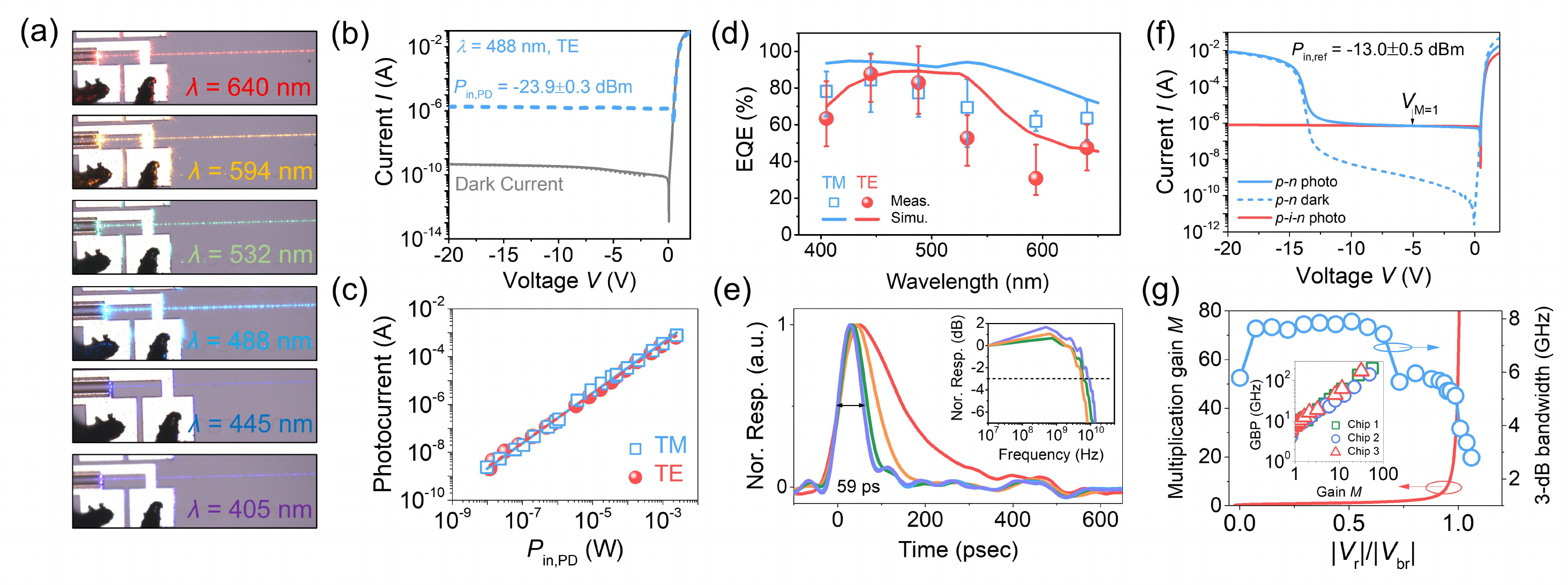}
\caption{Characterization of the SiN-on-Si PDs (device length: 50 $\mathrm{\mu m}$). (a) Optical micrographs of the same device under test at different wavelengths. (b-e) Characterization of PIN devices ($l$ = 50$\mathrm{\mu m}$, measured $W_{gn}$ = $200 \pm 20$ nm, $t$ = $120 \pm 10$ nm, and
$W_{gap} = 190 \pm 10$ nm). (b) Current-voltage ($I-V$) characteristics without (dark current) and with $P_{in,PD}$ of $-23.9\pm 0.3$ dBm. (c) $I_{eph}$ as a function of $P_{in,PD}$ from $\sim 10^{-9}$ to $10^{-3}$ W. (d) Measured EQE at -2 V agrees well with the simulation. Three devices from 3 chips far apart ($>10$ cm) on the wafer were measured. The error bars show the maximum and minimum quantities measured for the 3 devices, and the data point shows the average. (e) Normalized impulse response of the device at different reverse biases. Inset: normalized frequency response extracted from fast Fourier transform. Legend: red, 0 V; yellow, -2 V; green, -5 V; blue, -10V; and violet, -20 V. The data in (b-c) and (e) came from one of the devices in (d). (f-g) Characterization of PN junction APDs. (f) $I-V$ characteristics without and with $P_{in,ref} = -13.0 \pm 0.5$ dBm at $\lambda = 405$ nm, before coupling into the SiN waveguide. The data for the PIN device came from one of the devices in (d). (g) The avalanche multiplication gain ($M$) and 3-dB OE bandwidth as a function of $|V_r|/|V_{br}|$. Inset shows the gain-bandwidth product (GBP) of devices from 3 dies on the wafer.}\label{fig:fig2}
\end{figure*}

Figure {\ref{fig:fig2}}(a) shows optical micrographs of the same example device under test at different wavelengths. The images have not been post-processed. In our design, the input facet of the Si mesa was at the middle of the chip ($L/2$, Fig. \ref{fig:fig1}(d)), and the input power to the PD, $P_{in, PD}$, was thus taken as $(P_{in, ref}+P_{out, ref})/2$ using a reference waveguide with the same width $W_{gw}$, where the $P_{in, ref}$ and $P_{out, ref}$ are the measured optical powers (in dBm) into and out of the waveguide, respectively. 

\subsection*{PIN Photodiodes} 

First, we report the PIN PDs with $l = 50$ $\mathrm{\mu m}$, and $W_{gn} = 200 \pm 20$ nm, $t = 120 \pm 10$ nm, and $W_{gap} = 190 \pm 10$ nm as measured from cross-sectional TEM images, and TE polarized input unless otherwise stated Fig. {\ref{fig:fig2}}(b-e)). Fig. \ref{fig:fig2}(b) shows the measured current-voltage ($I-V$) characteristics of a device at $\lambda = 488$ nm. Measurements from $>20$ devices distributed uniformly on the wafer showed that the dark currents were, on average, $144 \pm 42$ pA and $266 \pm 65$ pA at reverse biases, $V_r$, of -5 and -15 V. No observable avalanche breakdown was found up to $-20$ V. At $\lambda = 488$ nm, a photocurrent of $\sim 1.35 ~\mathrm{\mu A}$ was observed at a $P_{in,PD}$ of $-23.9 \pm 0.3$ dBm at -2 V and was maintained throughout the $V_r$ range. The photocurrent included the absorption of the stray portion of $P_{in, ref}$ that was not coupled into the waveguide, which was eliminated by fiber displacement measurements (see Methods). We measured the effective photocurrent, $I_{eph}$, due to the light coupled into the waveguide without the stray light (as defined in Methods, Eq. \ref{eq:S1}) over six orders of magnitude of $P_{in,PD}$ ($\sim 10^{-9}-10^{-3}$ W) and found a good linearity in the photocurrent response (Fig. \ref{fig:fig2}(c)). The measurements of a PN device are included in the Supplementary Information (Fig. S3), and they also showed good $I_{eph}-P_{in,PD}$ linearity.

The EQE was calculated as $\mathcal{R} hc/(\lambda q)$, where $\mathcal{R} = I_{eph}/P_{in,PD}$ is the responsivity and $h$, $c$, $q$ denote Planck's constant, the speed of light and the elementary charge, respectively. Figure \ref{fig:fig2}(d) shows the measured EQE at $-2$ V. The measured EQE qualitatively agrees with but is lower than the simulated results, which assume an ideal internal quantum efficiency and lossless SiN waveguides and transitions (see Supplementary Information). The EQE was $> 60 \%$ for a spectral span of $\sim 135$ nm (400-535 nm) and $> 230$ nm (400-640 nm) for the transverse electric (TE) and TM polarized mode, respectively. For $\lambda \gtrsim 550$ nm, although the mode approached cut off at $W_{gn} = 200$ nm (Supplementary Information, Fig. S5(c)), the EQE benefits from the absorption of scattered light (Supplementary Information, Fig. S8). The measured EQE is limited by a number of factors, such as the internal quantum efficiency, mode mismatch loss between the input SiN waveguide and the SiN-on-Si region, $W_{gn}$ variations, device length, scattering and back-reflection at the Si mesa interface, and excitation of high order modes in the routing waveguide at short wavelengths (e.g., near 405 nm) (Supplementary Information). Nevertheless, Fig. S7 (Supplementary Information) shows a device length of $l = 50 ~\mathrm{\mu m}$ is sufficient to saturate the EQE for $\lambda > 445$ nm. We verified that longer devices (up to $l = 500 ~\mathrm{\mu m}$) showed an increase of $< 30\%$ in $\mathcal{R}$  (Supplementary Information, Fig. S9). Increasing $l$ or narrowing $W_{gn}$ can increase the EQE for the TE polarization at $\lambda <450$ nm (Supplementary Information, Fig. S10). Simulations show that the EQE is insensitive to the Si width, Si thickness, and $W_{gap}$ near our experimental value of $W_{gap} = 190$ nm (Supplementary Information, Fig. S11 and S12). $W_{gap} \in [110, 220]$ nm maximizes the EQE for both polarizations.

We extracted the PD OE bandwidth by inputting short pulses (full-width-at-half-maximum (FWHM) pulse width $<50$ ps) at a wavelength of 405 nm into the device and measuring the output voltage  in real-time (see Methods). Figure {\ref{fig:fig2}}(e) displays the measured impulse response of a device at different $V_r$ (the line labels are in the caption). The FWHM decrease with increasing $|V_r|$ can be attributed to the reduced carrier transit time due to an electric field in the depletion region, as well as the reduced junction capacitance due to a wider depletion width. The frequency response was obtained by applying the Fourier transform on the impulse response, as shown in the inset of Fig. \ref{fig:fig2}(e). The OE 3-dB bandwidth was $4.4 \pm 1.1$ and $8.6 \pm 1.0$ GHz at -2 and -20 V, respectively. The measured PIN junction capacitance and contact resistance (Supplementary Information, Figs. S13, 14) suggested that the measured bandwidth was limited by the carrier transit time at low bias ($|V_r| <10$ V), and the laser pulse width and instrumentation at high bias ($|V_r| >10$ V) (Supplementary Information, Fig. S16). 

\subsection*{PN Avalanche Photodiodes} 

The PN junction PDs could also function as avalanche photodiodes (APDs). Figure \ref{fig:fig2}(f) shows the I-V characteristics of such a device ($l = 50$ $\mathrm{\mu m}$ and measured $W_{gn} = 200 \pm 20$ nm, $t = 120 \pm 10$ nm, $W_{gap} = 190 \pm 10$ nm). Avalanche multiplication was significant beyond $V_r \sim -14$ V. An $I_{ph}-V$ characteristic of an identical PIN device with same $P_{in, ref}$ was used as a reference to determine the unity gain point, since $I_{ph}$ for the PIN device has been verified without avalanche gain up to -20 V (Fig. \ref{fig:fig2}(b)). The unity gain voltage ($V_{M=1}$) was determined as the intersection point of the $I_{ph}-V$ curves between the PN and PIN devices. Measurements from 3 chips across the wafer showed $V_{M=1} =-4.5 \pm 0.9$ V for $l = 50 ~\mathrm{\mu m}$ devices. We then found an avalanche multiplication gain, $M$, of $46 \pm 14$ and a corresponding gain-bandwidth product (GBP) of $173 \pm 30$ GHz at the avalanche breakdown voltage, $V_{br}$ (Fig. \ref{fig:fig2}(g)). Details on the determination of $M$, GBP, and $V_{br}$ are in the Methods section and Supplementary Information, as well as the performance of APDs with  $l = 100 ~\mathrm{\mu m}$.

\subsection*{A tunable microring integrated with a PD} 

\begin{figure}
\includegraphics[width=\columnwidth]{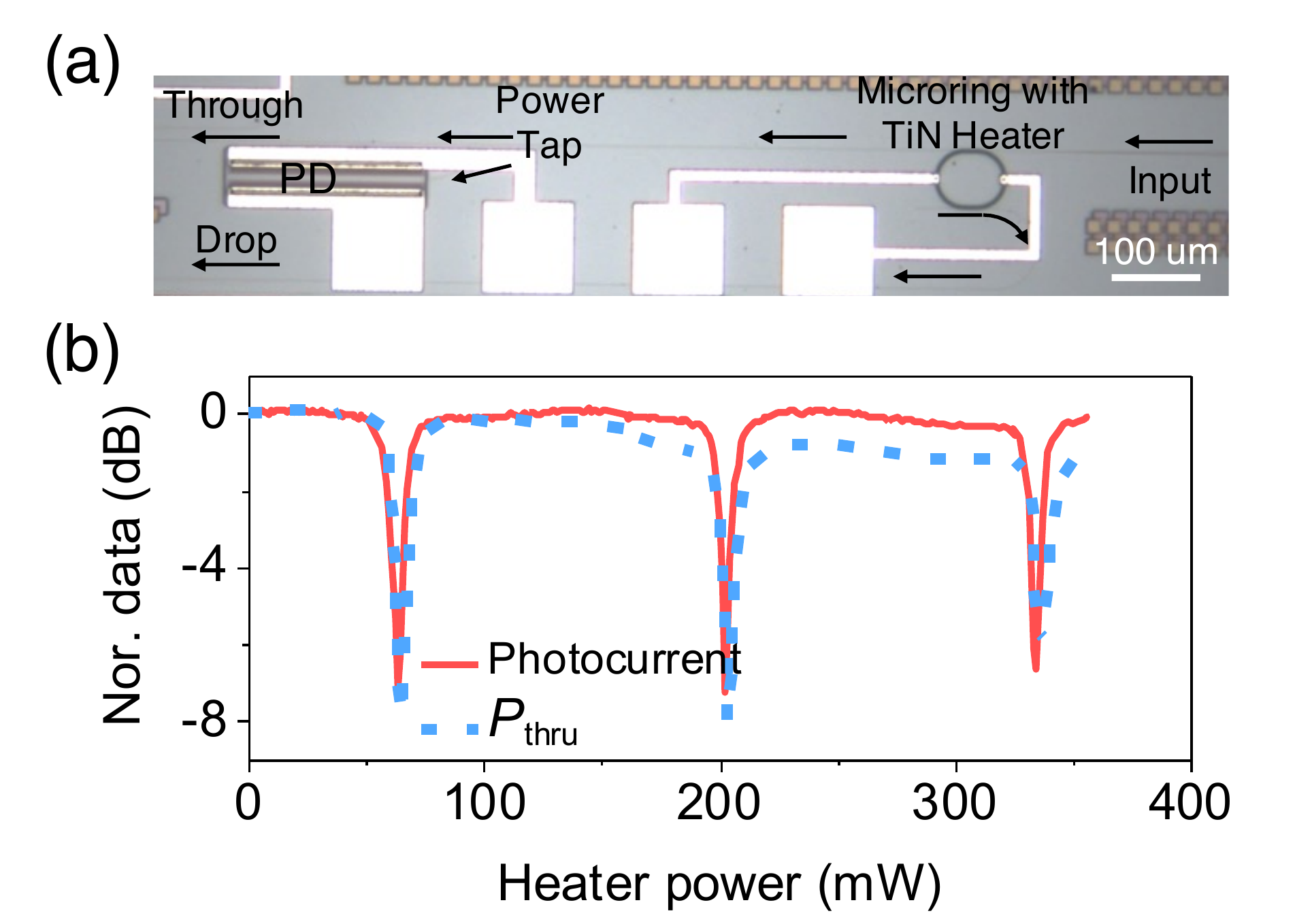}
\caption{On-chip characterization of a racetrack microring filter using an integrated PD.  (a) An optical micrograph of the PIC consisting of a thermally tunable microring with a PD connected to the power tap  at the through output port. (c) Normalized photocurrent and through port transmission, $P_{thru}$, as a function of the TiN heater power.}\label{fig3}
\end{figure}

Finally, as a proof-of-concept PIC demonstration, we integrated a PIN PD at the tap of the through port of a tunable SiN racetrack microring filter.  Fig. \ref{fig3}(a) shows an optical micrograph of the PIC. The TiN heater above the ring was used for thermal tuning, and the microring did not use the suspended heater structure in \cite{YongOE2022} which required clearance around the waveguide. We simultaneously measured the output power at the through port and the photocurrent while tuning the ring for an input wavelength of 514 nm. Details of the measurement are described in the Methods section. Figure \ref{fig3}(b) depicts a good match between the normalized through port transmission, $P_{thru}$, and photocurrent, demonstrating the feasibility of the PD for on-chip power monitoring. The slight increase in the insertion loss observed at the through port with the increase of the heater power was likely due to a  drift of the output coupling, since the insertion loss increase was not observed by the on-chip PD. To our knowledge, this is the first visible spectrum photonic circuit with a monolithically integrated photodetector.

\section{Discussion and Conclusion}

Our monolithically integrated SiN-on-Si waveguide PDs are unique and represent a new state of the art. Compared to the Si PD end-fired coupled to a SiN waveguide in \cite{YanikgonulNATCOMM2021}, which was characterized only at 685 nm, our SiN-on-Si PDs exhibited a higher EQE across a significantly wider wavelength span while maintaining a similar GBP ($173 \pm 30$ GHz in this work vs. $234 \pm 25$ GHz in \cite{YanikgonulNATCOMM2021}) and wafer-level performance uniformity (up to $\pm\sim30\%$ variation on dark current, EQE, $M$, OE bandwidth and GBP). The PDs reported here also have superior EQE and OE bandwidth compared to the heterogeneously integrated PDs in \cite{CuyversOL2022} and \cite{GherabliARXIV2021}. Si metal-semiconductor-metal (MSM) PDs integrated on SiN waveguides as in \cite{YuanPTL2006} and \cite{DeVitaARXIV2022} have $\sim$kHz OE bandwidths and are significantly slower than the PN and PIN devices demonstrated here. A comparison table is available in the Supplementary Information.

In summary, we demonstrated the first PDs monolithically integrated into a visible spectrum Si PIC platform. The SiN-on-Si PDs achieved broadband high-efficiency photon detection across the entire visible spectrum. Due to a leaky-wave design of a tapered SiN waveguide over a Si mesa, an EQE $> 60 \%$ was achieved over a wide wavelength span from 400 to 640 nm (for TM polarized light). The PDs had a 3-dB OE bandwidth up to 9 GHz and could also be used as APDs. Compared to other waveguide-integrated visible and NIR PDs, the device has the widest operating wavelength range while exhibiting similar dark current and APD gain-bandwidth product. The monolithically integrated SiN-on-Si PDs open the path toward sophisticated NIR and visible PICs. Future work includes the integration of PD arrays with an additional well doping opposite to the substrate doping type to avoid current leakage among devices and the investigation of the device operation as single-photon APDs (SPADs).

\section{Methods}

\subsection*{Fabrication} 
The integrated photonic platform was fabricated on 200-mm Si substrates at Advanced Micro Foundry. The fabrication begins with forming the PDs. First, doped regions were formed by ion implantation followed by rapid thermal annealing, and then mesas were etched. Next, the SiO$_2$ bottom cladding was deposited.  Two SiN waveguide  layers  were  then formed  by  plasma  enhanced  chemical  vapor  deposition  (PECVD), patterned by ArF  deep  ultraviolet lithography and  reactive  ion  etching. CMP  was  used  for  layer  planarization.  The metal  and  heater  layers  were then defined followed by deep trench and undercut etching for suspended structures and edge coupler facets.

Ellipsometry measurements show the SiN and cladding SiO$_2$ had refractive indices of $\sim 1.82$ and $\sim 1.46$, respectively, at $\lambda = 488$ nm.

\subsection*{DC characterization setup} 

For the responsivity and EQE measurement, the fabricated photodiodes were characterized using a continuous-wave multi-wavelength source (Coherent OBIS Galaxy). Cleaved Nufern S405-XP single-mode fiber was used to edge-couple the light into and out of the chip. The input and output fibers were mounted on 5-axis piezo-controlled micro-manipulators for a precise alignment. A polarization controller (Thorlabs CPC900) was inserted into the input fiber path for the polarization control. Optical power was measured using a free-space wand-style power detector (Newport 918D-ST-SL, for $P_{in,ref}$ measurement) and a fiber-optic detector (Newport 818-SL-L, for $P_{out, ref}$ measurement), and was then read out by a power meter (Newport 2936-R). $I-V$ characteristics were obtained using a Keysight B2912A Precision Source/Measure Unit (SMU) (source voltage and measure current) via two DC electrical probes (MPI MP40 Micro-positioner, tungsten probe tip) placed on device anode and cathode contact pads. The voltage sweep ranged from forward (+2 V) to reverse (-20 V) bias in steps of 0.02 V.  We verified that the photocurrent from ambient light was negligible during the dark current sweeps. Each photocurrent sweep was carried out at the respective maximum photocurrents by performing careful fiber
alignment with the input SiN waveguides. 

The photocurrent also included the absorption of stray light from $P_{in,ref}$  that was not coupled into the waveguide. To estimate this contribution, the fiber was horizontally displaced $\sim 3$ $\mathrm{\mu}$m  from the optimal fiber-waveguide alignment to eliminate the waveguide-coupled light absorption (mode field diameter $\sim 3$ $\mathrm{\mu}$m for Nufern S405-XP). The corresponding photocurrent ($I_{disp}$)  was then multiplied by a
term ($1-10^{-\eta/10}$) to factor out the photocurrent due to the light that was coupled into the waveguide at the optimal alignment, where $\eta $ is the single-facet coupling loss of the edge coupler (in dB). Here it is assumed that the 3 $\mathrm{\mu}$m fiber displacement resulted in no change of the absorption of the uncoupled light into the PD. The obtained value was then subtracted from the $I_{ph}$ to get the estimated effective photocurrent
($I_{eph}$) due to the  light coupled into the waveguide. The above description can be expressed by 
\begin{equation}\label{eq:S1}
I_{eph} = I_{ph} - I_{disp} \left(1 - 10^{-\frac{\eta}{10}} \right).
\end{equation}
In future designs, the input and output waveguides can be displaced from each other to eliminate the influence of stray light.

\subsection*{OE bandwidth characterization setup}
 
To measure the PD OE bandwidth, a train of optical pulses ($\lambda = 405$ nm, full-width-half-maximum (FWHM) $<50$ ps) at a repetition rate of 10 MHz from a picosecond laser diode (PicoQuant LDH-D-C-405 laser head with PDL 800-D driver) was coupled into the PDs. The resultant electrical impulse response from a PD was captured on a 13-GHz real-time oscilloscope (Keysight Infiniium UXR0134A) connected  to an RF probe (GS configuration, GGB Industries Picoprobe 40A), bias-tee (API Technologies, Inmet 8810KMF1-40, 25 V, up to 40 GHz) and a 40-GHz RF cable. We did not target a specific input polarization of the light for this measurement. The maximum optical power from the laser was 0.72 mW, and no nonlinear effects were observed.

\subsection*{Capacitance and contact resistance measurements}

The device capacitance was measured using an impedance analyzer (Keysight E4990A, with a 42941A impedance probe) via frequency sweeps from 20 Hz to 1 MHz. The frequency range is commonly employed for the capacitance measurement of semiconductor PN junctions \cite{LuciaEJP1993}. The same electrical probes used for the $I-V$ measurements were connected for device probing. Standard equipment calibration, followed by fixture (i.e., electrical cables and probes) compensation (at ``OPEN'') was performed before the measurement. The  capacitance of devices at different lengths and reverse bias voltages were measured to extract the per-length junction and parasitic capacitance. The measurements are shown in the Supplementary Information. The parasitic capacitance is found to be $70 \pm 21$ fF and $ 55 \pm 17$ fF for PIN and PN  devices, respectively, throughout the measured reverse biases. 

The contact resistance was extracted from the device $I-V$  characteristics at forward bias. In an ideal forward-biased PN junction, the current increases exponentially with voltage. In practice, the current switches to a linear increase beyond a certain forward-bias voltage, when an external resistance (e.g., contact resistance) becomes more significant than the junction forward resistance.

From the extracted capacitance and resistance, and accounting for 50 $\Omega$ load resistance from the measurement apparatus, the calculated $RC$-limited 3-dB OE bandwidth for the 50-{\textmu}m long devices approaches 30 GHz at $|V_r| \sim 10$ V for both PN and PIN configurations (see Supplementary Information), significantly higher than the measured results, which were limited by the laser pulse width and the instrumentation.

\subsection*{Definition of APD Parameters}

The avalanche multiplication gain, $M$, is given by 
\begin{equation}
M(V_r) = \frac{I_{ph}(V_r) - I_{dark}(V_r)}{I_{ph}(V_{M=1}) - I_{dark}(V_{M=1})},
\end{equation}
where $I_{dark}$ is the dark current, $V_r$ is the reverse bias voltage, and $V_{M=1}$ is the bias voltage at $M=1$. The avalanche breakdown voltages $V_{br}$ can be extracted from \cite{MaesSSE1990}
\begin{equation}\label{eq:VrVbr}
\left( \frac{V_r}{V_{br}}\right)^n = 1- \frac{1}{M(V_r)},
\end{equation}
via linear interpolation, where $n$ is an empirical parameter. $V_{br}$ was found to be $-13.3 \pm 0.9$V and $-13.8 \pm 1.1$V, respectively, for 50 and 100 {\textmu}m-long devices (see Supplementary Information). As $V_{br}$ varied among devices, we used $|V_r|/V_{br}|$ to compare $M$ and the 3-dB OE bandwidth of different devices. The bandwidth measurement was identical to that for Fig. 2(e), except we held $P_{in, ref} = -13.0 \pm 0.5$ dBm, the same as that in the avalanche gain measurement. As $V_r$ approached $V_{br}$,  $M$ substantially
increased, while the 3-dB bandwidth decreased. At $V_r = V_{br}$, $M = 46 \pm 14$ and $29 \pm 9$, respectively, for the 50- and 100-{\textmu}m long devices. Correspondingly, the 3-dB bandwidth reduced from $7.6 \pm 0.3$ to $4.0 \pm1.4$ and $3.7 \pm1.6$ GHz. The influence of the input pulse width (estimated 3-dB OE bandwidth $\sim 9$ GHz) and the oscilloscope bandwidth (13 GHz) on the device 3-dB bandwidth can
thus be considered negligible at $V_{br}$. Similar bandwidth reductions have been observed in other works \cite{YanikgonulNATCOMM2021, HuangOptica2016, BenedikovicOptica2020, ChenOE2015,ZhuJSTQE2018}  and can be explained by an increased avalanche build-up time at higher $M$. We further calculated gain-bandwidth product (GBP) up to $M(V_{br})$. At $V_r = V_{br}$, GBP = 173{\textpm}30 and 99{\textpm}15 GHz, respectively,
or 50- and 100-{\textmu}m long devices. The values are reasonable compared to \cite{YanikgonulNATCOMM2021, YuanJS2022}, indicating good APD performance. The $M$, 3-dB OE bandwidth and GBP results for the 100-{\textmu}m long devices can be found in the Supplementary Information (Fig. S17).

The avalanche breakdown voltage $V_{br}$  was determined using Eq. \ref{eq:VrVbr}, which is equivalent to
\begin{equation}
\ln \left(1 - \frac{1}{M(V_r)} \right) %= n \cdot \ln \left( \frac{|V_r|}{|V_{br}|}\right) 
= n \ln (|V_r| ) - n  \ln (|V_{br}|). 
\end{equation}
Therefore, performing a linear fitting and extrapolation between $ \ln \left(1 - \frac{1}{M(V_r)} \right)$ and $\ln(|V_r|)$ leads to the extraction of $n$ and $n \ln(|V_{br}| )$, via its slope, $a$, and $y$-intercept, $b$, respectively. $V_{br}$ can thus be calculated using $|V_{br}| = \exp(-b/a)$.  The Supplementary Information (Fig. S18) shows an example plot of the
fitting for a 50-{\textmu}m long PN device, resulting in a $V_{br} = -13.1$ V.  $V_{br}$  of other device lengths were similarly determined using this approach.

\subsection*{Wavelength-tunable microring filter measurement}

The photonic integrated circuit comprises a SiN racetrack microring resonator (radius = 28 {\textmu}m, straight arm length = 12 {\textmu}m) with a TiN resistive heater above, and a PIN photodiode (200 {\textmu}m-long) coupled from the bus waveguide (at the through port) of the ring via a directional coupler. The measurement setup is similar to the DC characterization setup described above, except we used  two additional electrical probes (MPI MP40) to power the heater via the other channel of the Keysight B2912A SMU. TE-polarized light at $\lambda  = 514$ nm from the Coherent OBIS Galaxy was coupled into the bus waveguide and both the photocurrent from the PD and the transmitted optical power at the through port, $P_{thru}$, were recorded from the SMU via script control, as a function of the applied electrical power to the heater. 

\subsection*{Data availability} 
Data underlying the results presented in this paper are available from the authors upon reasonable request.

\subsection*{Acknowledgments} 

The authors thank J. Groth for assistance on data analysis, A. Stalmashonak for assistance in the measurements. The authors are  grateful for the loan of the picosecond laser from PicoQuant GmbH.

%\bibliography{references2}


\pagebreak
\widetext
\begin{center}
\textbf{\large Supplementary Information}
\end{center}

%%%%%%%%%% Merge with supplemental materials %%%%%%%%%%
%%%%%%%%%% Prefix a "S" to all equations, figures, tables and reset the counter %%%%%%%%%%
\setcounter{equation}{0}
\setcounter{figure}{0}
\setcounter{table}{0}
\setcounter{page}{1}
\makeatletter
\renewcommand{\theequation}{S\arabic{equation}}
\renewcommand{\thefigure}{S\arabic{figure}}
\renewcommand{\bibnumfmt}[1]{[S#1]}
\renewcommand{\citenumfont}[1]{S#1}
%%%%%%%%%% Prefix a "S" to all equations, figures, tables and reset the counter %%%%%%%%%%

\section{Via separation and Si mesa height}

\begin{figure}[!ht]
\centering  \includegraphics[width = 0.9\textwidth]{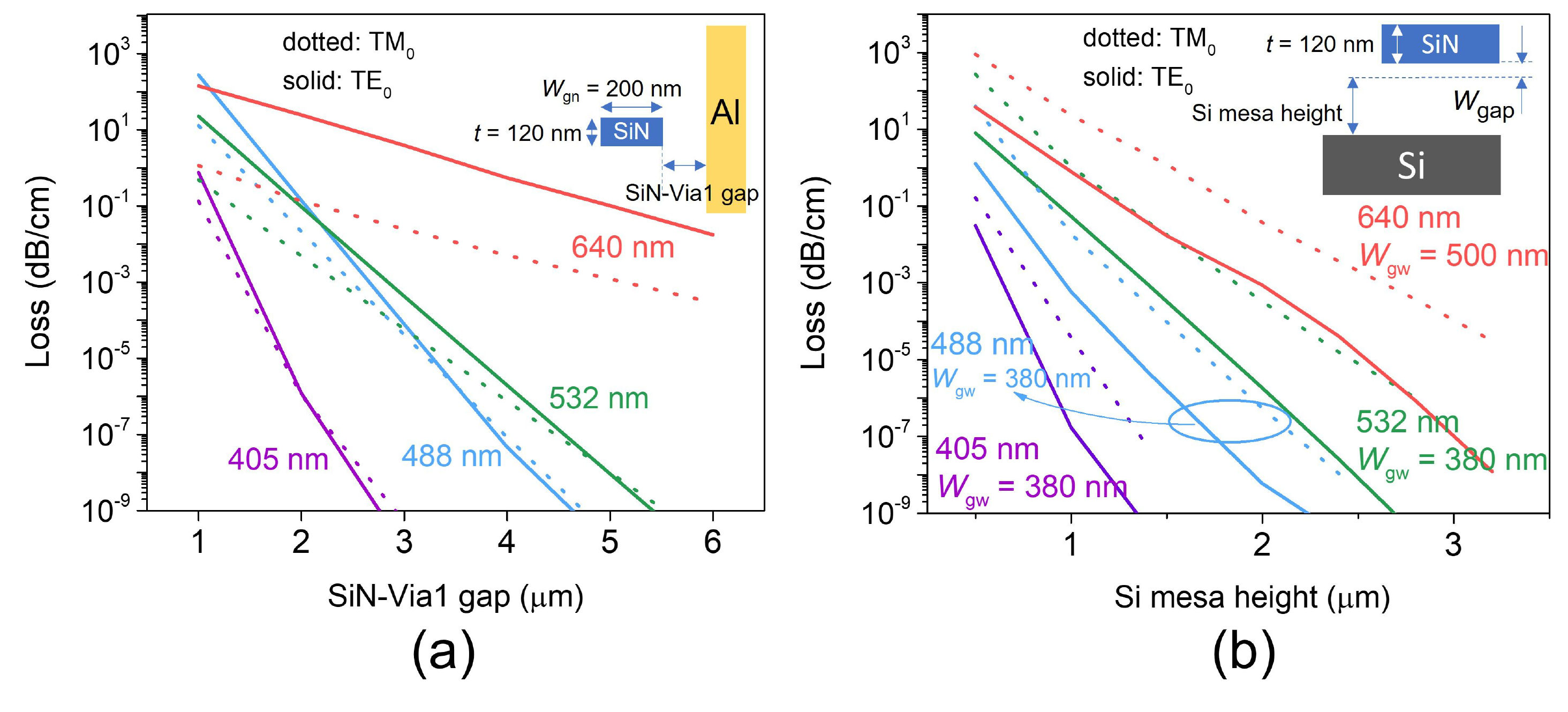} 
\caption{Calculated SiN waveguide loss as a function of (a) SiN-Via1 gap and (b) Si mesa height at wavelengths of 405, 488, 532 and 640 nm} \label{fig:S1}
\end{figure}

We computed the SiN waveguide propagation loss using a finite difference eigenmode (FDE) solver (Lumerical MODE Solutions) to set the Al via separation and the Si mesa height in the PD design. Figure {\ref{fig:S1}}(a) shows the waveguide loss for the $\mathrm{TM}_0$ and $\mathrm{TE}_0$ modes at $\lambda = 405$, 488, 532 and 640 nm as a function of SiN-Via1 gap. The inset shows the schematic for the SiN waveguide and Via1 parameters used in the calculation. The Si substrate is not included to isolate the effect of via. The loss due to the Al via is $< 0.02$ dB/cm for all the wavelengths at a gap of 6 {\textmu}m, which is the minimum separation in the PDs presented.

Figure {\ref{fig:S1}}(b) shows the waveguide loss as a function of Si mesa height. Here, the waveguide widths ($W_{gw}$) are 380 nm at $\lambda$ = 405, 488 and 532 nm, and 500 nm at $\lambda$ = 640 nm. For $W_{gw}$ = 380 nm, the loss is $< 10^{-6} $ dB/cm at the mesa height of 2.85 $\mu$m for wavelengths up to 532 nm. For longer wavelengths, the loss can be less than $4 \times 10^{-4}$ dB/cm for $W_{gw}$ = 500 nm up to $\lambda$ = 640 nm. Therefore, the Si mesa height was set to 2.85 $\mu$m with $W_{gw}$ = 500 nm for a low propagation loss up to $\lambda$ = 640 nm, and we used $\lambda$ = 514 nm for the tunable microring design and measurement with $W_{gw}$ = 380 nm.

These calculations used a background index of 1.46 and metal boundary conditions. At the boundary, at $\lambda$ = 640 nm, the electric field magnitude was $10^{-7}$ of the maximum.

\section{Edge coupling and waveguide propagation loss}

\begin{figure}[!ht]
\centering  \includegraphics[width = 0.9 \textwidth]{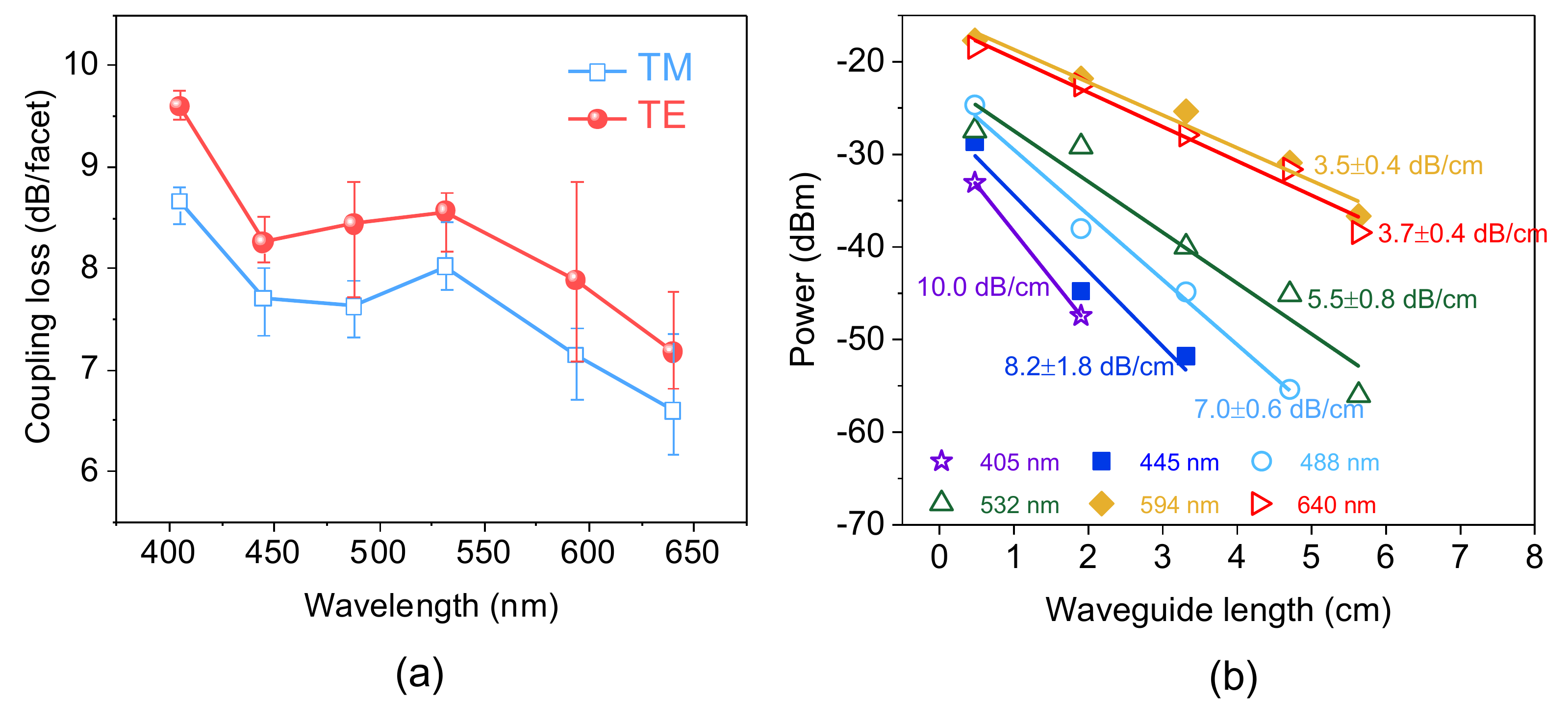} 
\caption{Measured (a) coupling loss of SiN tapered edge couplers (from 3 chips far apart on wafer) and (b) propagation loss of routing SiN waveguides (width = 520 nm) at TE polarized mode.}\label{fig:loss}
\end{figure}

Figure \ref{fig:loss} show the measured edge coupler and waveguide propagation loss. Cleaved single-mode fibers coupled light into and out of the respective test structures on chip \cite{SacherOE2019s, LinOE2021s}. The higher coupling loss at short wavelengths is due to the lower mode overlap between the fiber and on-chip edge coupler \cite{LinOE2021s}. The lower coupling loss for TM relative to TE is due to the reduced optical confinement of the fundamental TM mode in the edge coupler and correspondingly a better mode overlap with the fiber \cite{SacherOE2019s}. The measurement error is due to the variability in the input/output coupling  and  waveguide loss. We have also observed similar loss variability on the wafers in \cite{LinOE2021s}.

\section{PD linearity and junction difference}

\begin{figure}[!htb]
\centering  \includegraphics[width = 0.5 \textwidth]{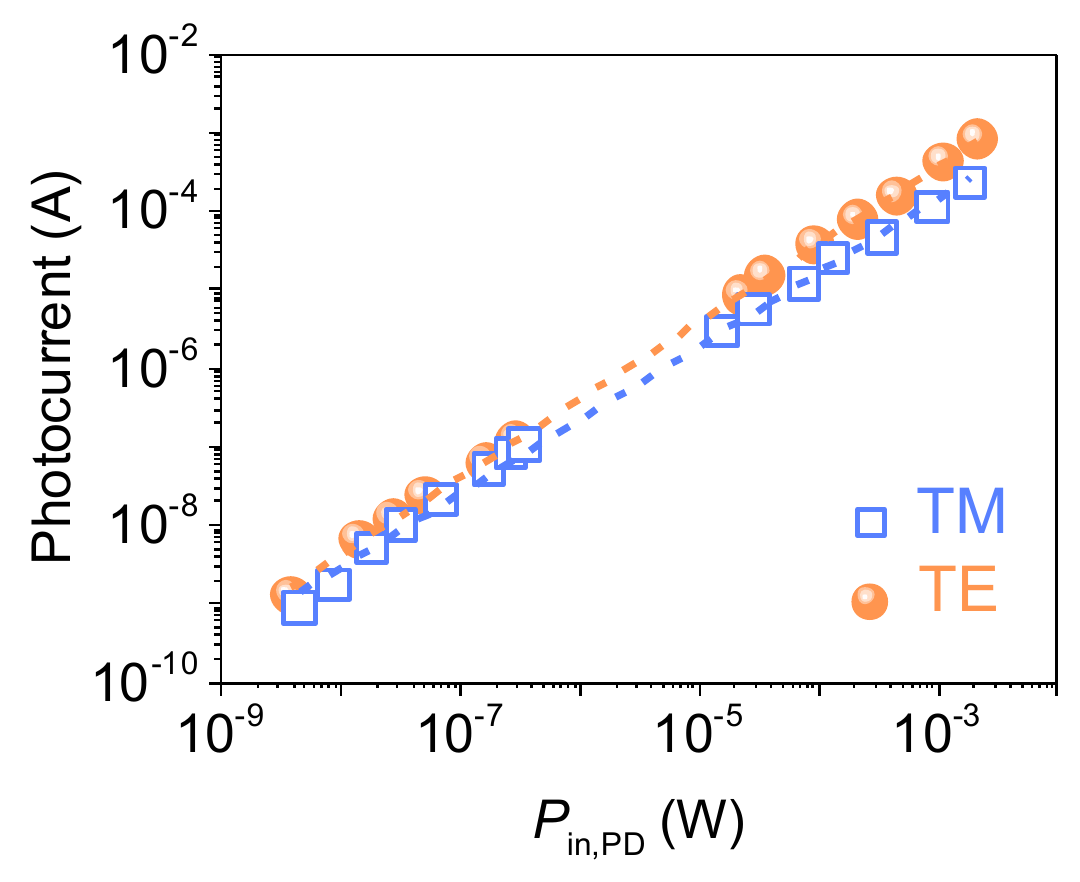} 
\caption{$I_{eph}$ for a 50-{\textmu}m long PN device as a function of $P_{in,PD}$ from $\sim 10^{-9}$ to $10^{-3}$ W at $\lambda = 488$ nm. A good linearity for both TE and TM polarizations was observed within this range of power. Responsivities: TE: $0.33 \pm 0.05$ A/W, TM: $0.30 \pm 0.05$ A/W.}\label{fig:S2}
\end{figure}

\begin{figure}[!htb]
\centering  \includegraphics[width=0.5 \textwidth]{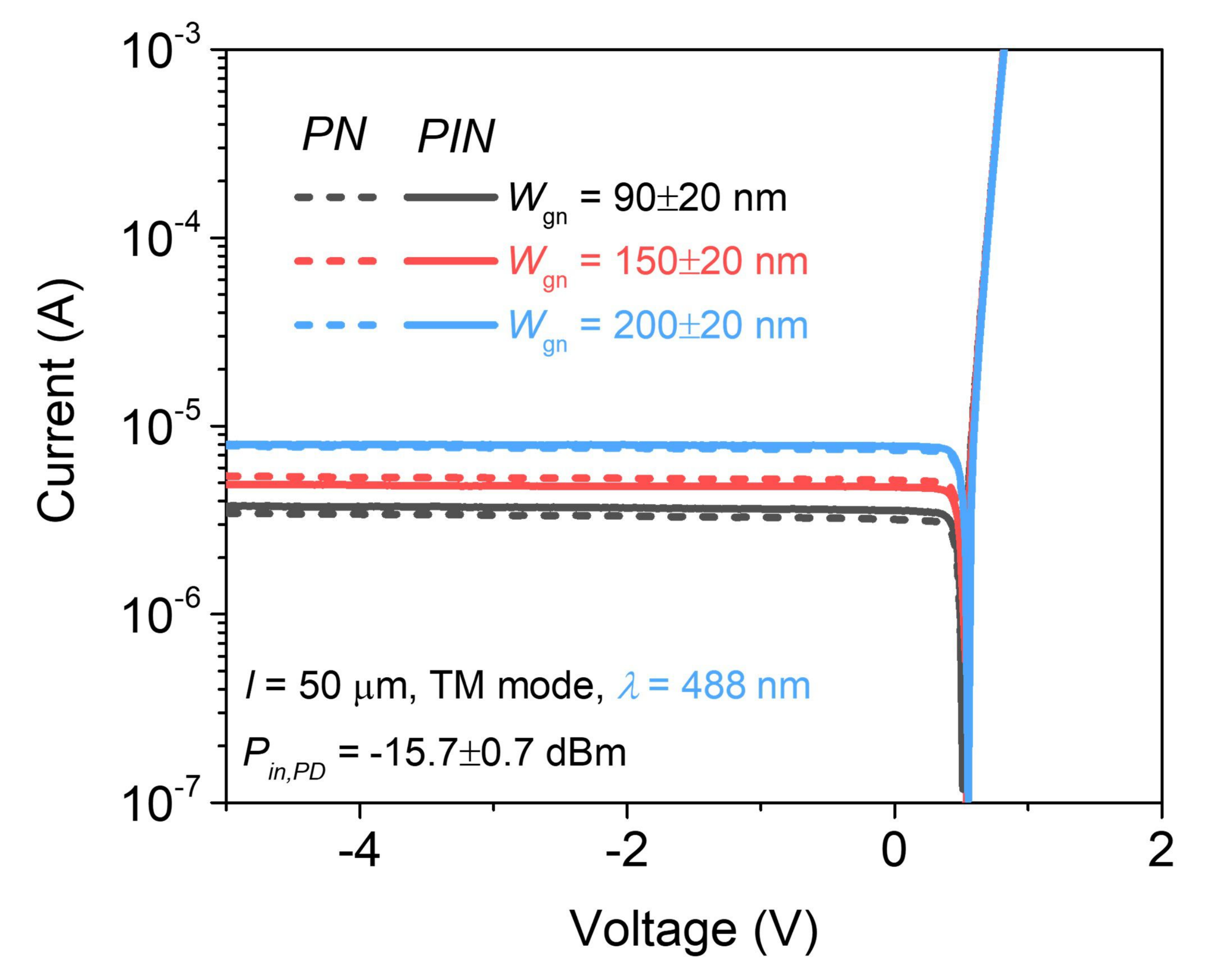} 
\caption{Measured photocurrents of PN and PIN PDs at different $W_{gn}$ (measured from TEM images) as a function of applied voltage at $\lambda$ = 488 nm.} \label{fig:S4add}
\end{figure}

Fig \ref{fig:S2} shows the effective photocurrent is a linear function of the input power of the 50-{\textmu}m long PN PD. Figure \ref{fig:S4add} shows that the observed photocurrents of the PN and PIN PDs are nearly identical.

\section{Calculation of $\eta_{mode}$} \label{sec:eta_mode}

\begin{figure}[!ht]
\centering  \includegraphics[width=0.8 \textwidth]{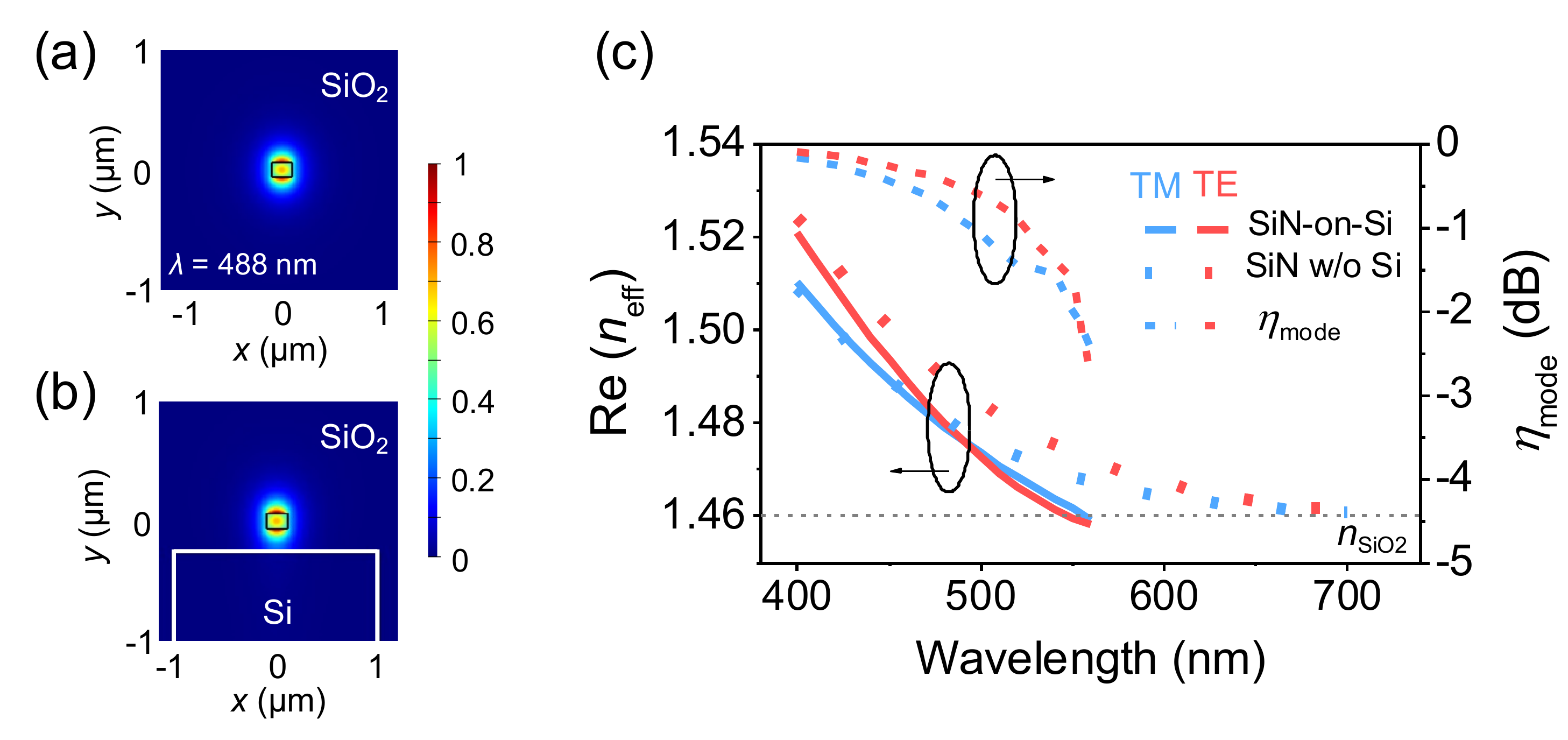} 
\caption{TM\textsubscript{0} mode profiles at $\lambda $ = 488 nm in the SiN waveguide ($W_{gn} = 200$ nm) (a) without and (b) with the Si mesa. (c) Simulated Re($n_{eff}$) of SiN without (as in (a)) and with (as in (b)) the Si underneath, as well as their corresponding mode mismatch loss ($\eta_{mode}$, in dB), as a function of wavelength.} \label{fig:S3}
\end{figure}

Figures \ref{fig:S3}(a) and (b) show, respectively, the TM$_0$ mode profiles at $\lambda $ = 488 nm in the narrowed part of the SiN waveguide ($W_{gn}$) and in the SiN-on-Si mesa region. The waveguide  dimensions used in the calculation are the average values from the TEM images $W_{gn} = 200$ nm, $t = 120$ nm, and $W_{gap} = 190$ nm). Figure \ref{fig:S3}(c) shows the computed mode mismatch loss ($\eta_{mode} $) determined from the corresponding mode overlap using a finite difference eigenmode (FDE) solver (Lumerical MODE Solutions). The refractive indices used for SiN and SiO$_2$  cladding were 1.82 and 1.46, respectively, for all wavelengths in the calculation (400-700 nm). The Si material properties were imported from the default database ``Si (Silicon) - Palik''. The width of Si was 2 {\textmu}m in the simulation and the structure was extended down beyond the simulation region to mimic bulk Si. Extending the Si width leads to a negligible change on $\eta_{mode}$. The 2-D simulation region on the cross-sectional plane of the waveguide spans 5 {\textmu}m in width and 4 {\textmu}m in height, with the SiN waveguide
sharing the same center point. Metal boundaries were used at the boundaries of the simulation region. Mode overlap calculation was performed between the eigenmode profiles  with only the SiN and additionally with the Si
underneath for $\eta_{mode} $. Due to the similarity of the mode profiles (Fig. \ref{fig:S3}(a) and (b)), the $\eta_{mode} $ is $< -2$ dB for $\lambda \in [400, 500]$  nm for both TE and TM polarized light (Fig. \ref{fig:S3}(c)). The computation was not extended to longer wavelengths since the real part of SiN-on-Si effective indices (Re($n_{eff}$)) dropped below the cladding index (1.46) at $\lambda > 550$  nm (Fig. \ref{fig:S3}(c)). The mode mismatch loss increases at longer wavelengths due to the reduced waveguide confinement factor \cite{AhnJLT2010}.

\section{Simulation of EQE}\label{sec:EQE}

\begin{figure}[!ht]
\centering  \includegraphics[width=0.6 \textwidth]{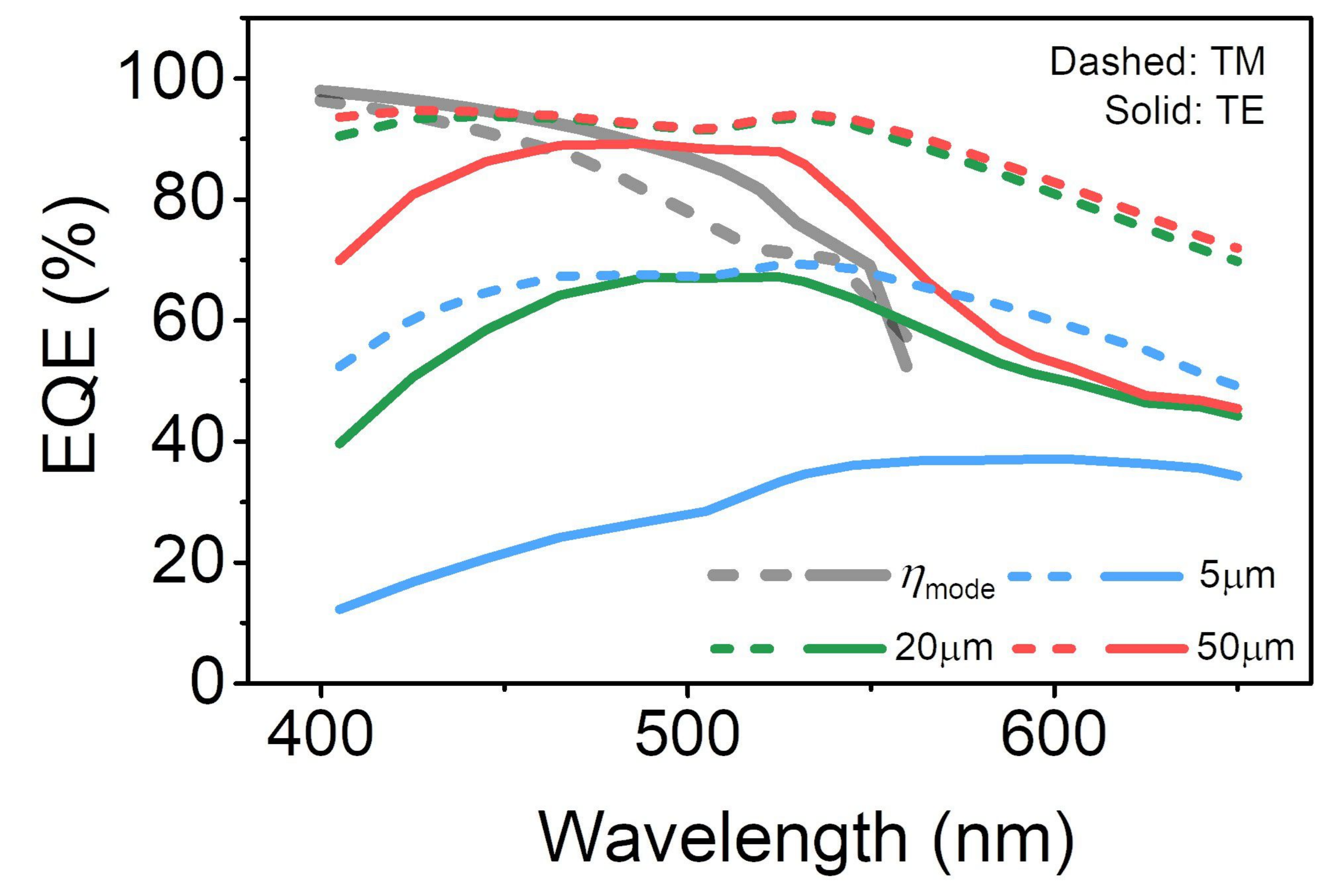} 
\caption{Simulated EQE vs. device length ($l$ , in Fig. 1(d), $l = 5, 20, 50$ {\textmu}m).  $\eta_{mode} $ (from Fig. \ref{fig:S3}(c)) is included for comparison.} \label{fig:S4}
\end{figure}

The simulation of the PD EQE was performed using the 3D finite-difference time-domain method (Lumerical FDTD). Material
property settings were identical to that in Section \ref{sec:eta_mode}. The Si mesa had a width of 2 {\textmu}m, and the Si thickness was 5 {\textmu}m for $\lambda \leq 532$ nm and 9 {\textmu}m for $\lambda > 532$ nm. The Si thickness leads to an error of $< 1\%$ in the total absorbed optical power for a 50 {\textmu}m long device.  Average waveguide dimensions $W_{gn} = 200$ nm, $t= 120$ nm, and $W_{gap} = 190$ nm) from the TEM images were used.  Perfectly matched layers (PMLs)  were used at the simulation region boundaries. To calculate the EQE, the total absorbed power in Si, given by the absorbed power density, $-\frac{1}{2}\omega\mathrm{Im}(\epsilon) |E|^2$, was integrated over the Si region and normalized to the input optical power. This calculation assumes a perfect internal quantum efficiency, i.e., all absorbed photons are converted to carriers collected by the electrodes. %{\textcolor{red}{This assumption is acceptable since an electric field for a high-efficiency carrier collection usually exists in the depletion region of a reverse-biased PN junction  \cite{LiuAPL2005, LiJJAP2015, ChatterjeeOL2019}.}} 
The EQE at device lengths ($l$, Fig. 1(b)) of 5, 10, 20, 30, and 50 {\textmu}m, from operating wavelengths of 400 to 700 nm were calculated. The cross-sectional area of the injected mode source was kept  constant at a width of 4 {\textmu}m and height of 3.8 {\textmu}m. For device lengths $< 50$ {\textmu}m, the simulation contained the 50 {\textmu}m-long device but the power density was integrated to the device length. These simulation settings were applied to all subsequent EQE calculations here. Fig. \ref{fig:S4} shows the calculated EQE at $l =$ 5, 20 and 50 {\textmu}m, where $\eta_{mode} $ (Fig. \ref{fig:S3}(c)) was also included for comparison. A device with $l$= 50 {\textmu}m approached the EQE limited by $\eta_{mode}$ between the input SiN waveguide and the SiN-on-Si region. The simulation results are generally higher than the measurements (Fig. 2(d)), since the simulations assume a perfect internal quantum efficiency and neglect losses. We observed negligible EQE difference between PN and PIN devices (Fig. {\ref{fig:S4add}}).

\section{ Absorption mechanism analysis}\label{sec:absorption}

\begin{figure}[ht]
\centering \includegraphics[width = 0.9 \textwidth]{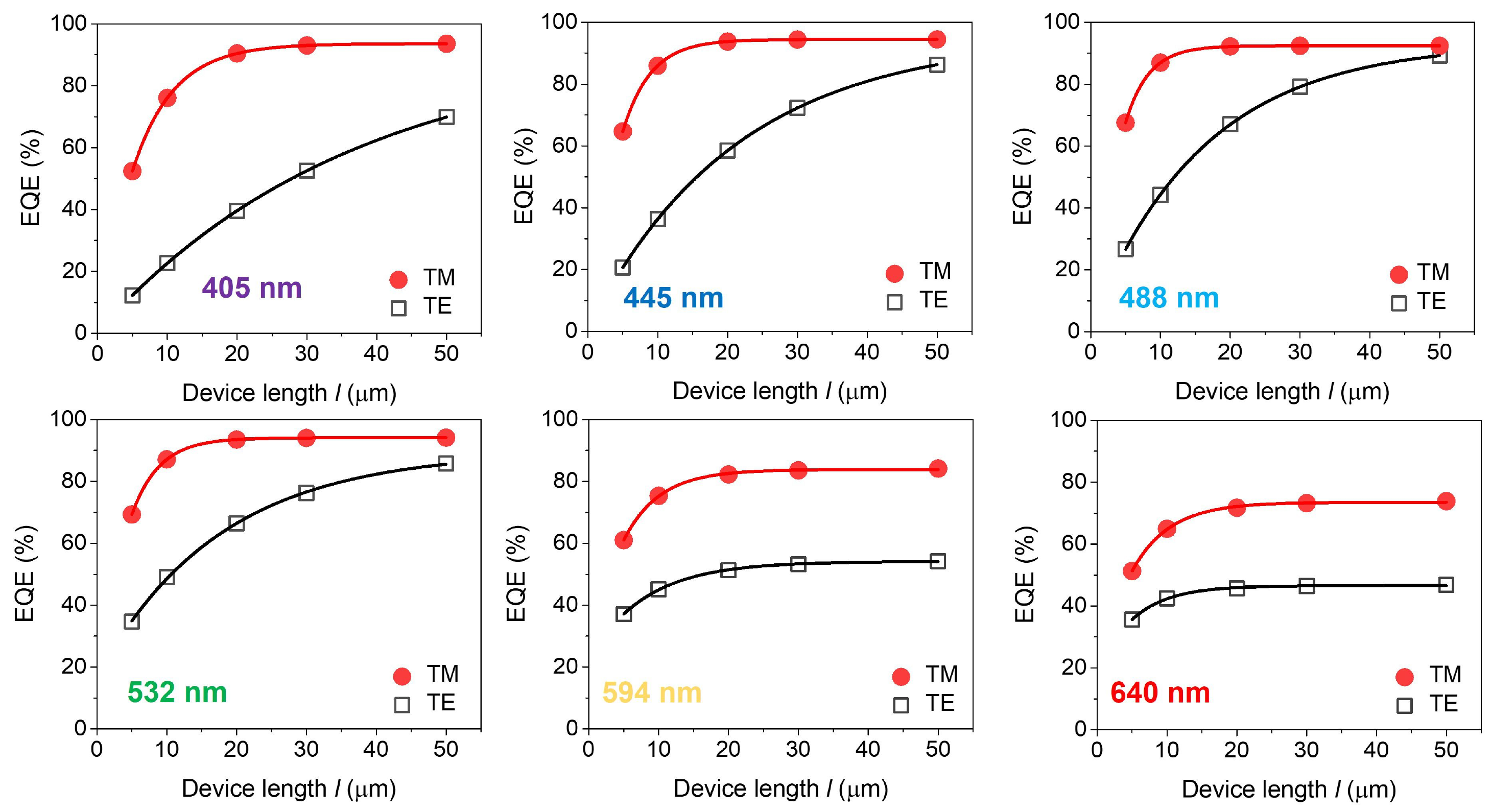}
\caption{ Simulated EQE as a function of device length $l$ at different  wavelengths. The data points are fitted with an exponential and the extracted parameters are tabulated in Table S1.}\label{fig:S5}
\end{figure}

\begin{figure}[ht]
\centering  \includegraphics[width=0.8 \textwidth]{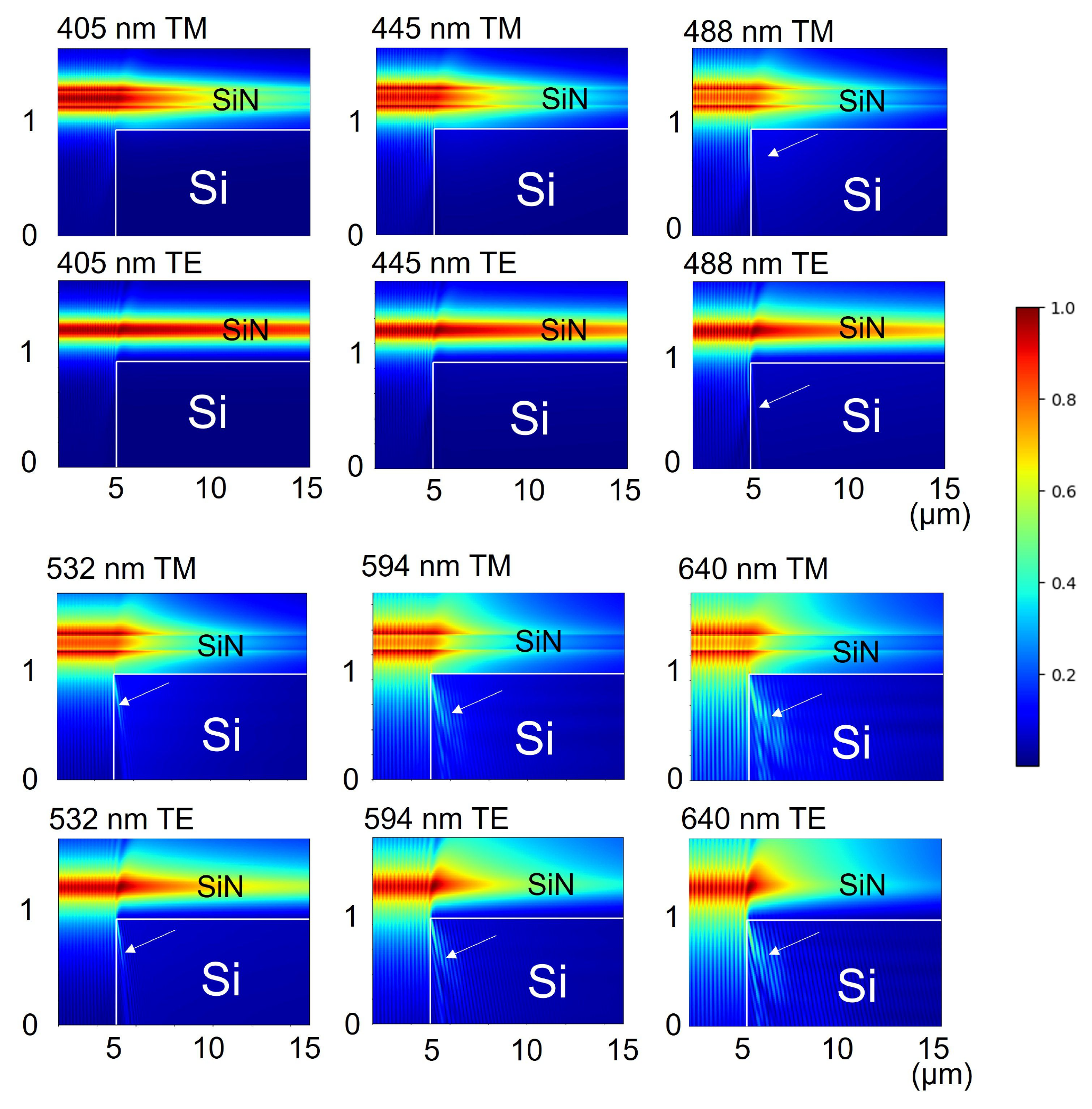}
\caption{ Cross-sectional $|E|$ profiles at different visible wavelengths. The cladding is SiO$_2$. Light scattering into Si is more significant at longer wavelengths since the input mode is less confined in the SiN waveguide. The arrows are not included in the figures at $\lambda$ = 405 and 445 nm due to weak scattering.}\label{fig:S6}
\end{figure}

We choose the 6 wavelengths (405, 445, 488, 532, 594 and 640 nm) used in the measurements to study the light absorption mechanisms in Si. Figure {\ref{fig:S5}} shows the simulated EQE as a function of device length $l$ at these wavelengths. First, we fitted the EQE to \cite{AhnAPL2009}
\begin{equation}\label{eq:S2}
\mathrm{EQE} = A \mathrm{e}^{-\alpha_{coupling} l} + C,
\end{equation}
where $A$ and $C$ are constants and $\alpha_{coupling}$ is the coupling efficiency into Si. $ A + C$ is the theoretical EQE as $ l \rightarrow 0$, and $C$ is the maximum achievable EQE as $l \rightarrow \infty$. The solid lines in Fig. \ref{fig:S5} are the fits and the fitting parameters are tabulated in Table \ref{table:EQE}. Generally, for $\lambda \geq 532$ nm where  $(A + C) > 0$ and $A+C$ increases with wavelength, a fraction of the launched light in the SiN waveguide is absorbed in the Si at the front facet of the mesa. The trend agrees  with the cross-sectional $|E|$ profiles in Fig. \ref{fig:S6}, where $|E|$  at the input Si mesa facet increases in amplitude with increasing wavelength (indicated by arrows). The TM polarized mode at longer wavelengths  is more strongly scattered into Si since the input Si facet represents a greater perturbation to the mode. The EQE for $\lambda \geq 532$ nm is due to a combined effect of the absorption of scattered light and the evanescent coupling between SiN and Si, while the EQE at the shorter wavelengths is mainly achieved by evanescent coupling.  $\alpha_{coupling}$  monotonically increases with wavelength for the TE polarization, but shows a peak at $\lambda = 488$ nm for the TM polarization. This is a result of the increasing mode overlap  with the Si in conjunction with the lower Si absorption as the wavelength increases. As the wavelength increases to $\sim 532$ nm, the TM mode is cut off first, leading to a reduction of $\alpha_{coupling}$. 

\begin{table}
\caption{Fitted parameters for EQE vs. $l$ (Eq. \ref{eq:S2}) at different visible wavelengths.} \label{table:EQE}
\begin{tabular}{|c|c|c|c|c|c|c|c|c|}
\hline
\multirow{2}{*}{Wavelength (nm)} & \multicolumn{2}{c|}{$A$} &\multicolumn{2}{c|}{$C$} & \multicolumn{2}{c|}{$A+C$} & \multicolumn{2}{c|}{\makecell{$\alpha_{coupling}$\\ ({\textmu}m\textsuperscript{-1})}}\\
\cline{2-9}
& TM & TE & TM & TE & TM & TE & TM & TE \\
\hline
 405 & -0.97 & -0.94 & 0.94 & 0.94 & -0.03 & -0.00 & 0.17 & 0.03\\
 445 & -1.04 & -0.94 & 0.94 & 0.95 & -0.10 & 0.01 & 0.25 & 0.05\\
488 & -1.13 & -0.91 & 0.92 & 0.93 & -0.22 & -0.02 & 0.30 & 0.06\\
532 & -0.88 & -0.73 & 0.94 & 0.90 &  0.06 & 0.17 & 0.25 & 0.06\\
594 & -0.60 & -0.31 & 0.83 & 0.54 & 0.23 & 0.23 & 0.19 & 0.12\\
 640 & -0.56 & -0.28 & 0.73 & 0.47 & 0.17 & 0.18 & 0.19 & 0.19\\
\hline 
\end{tabular}
\end{table}

\section{Responsivity vs. device length}

\begin{figure}[ht]
\centering \includegraphics[width = 0.3 \textwidth]{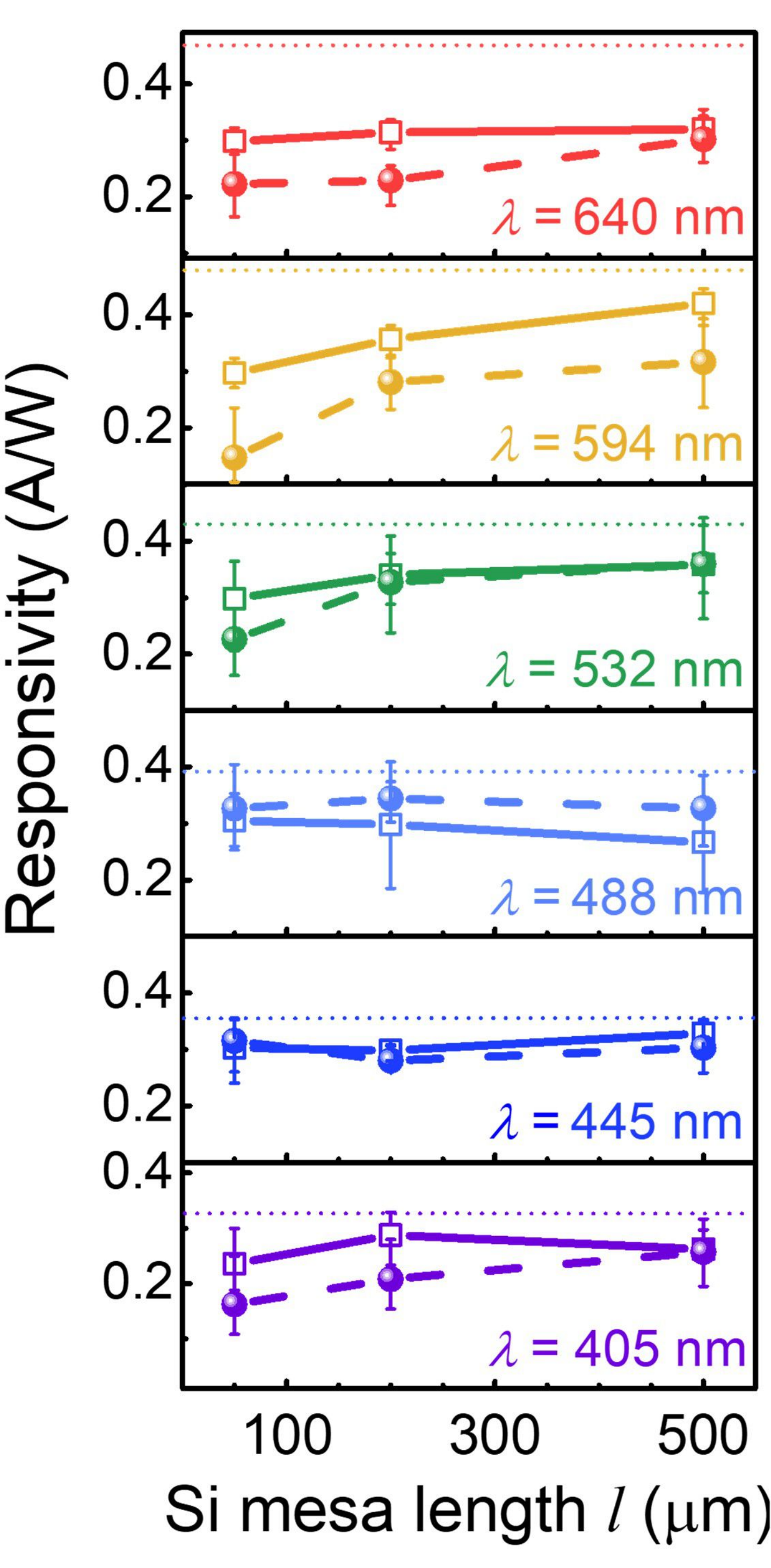}
\caption{Measured responsivity of PIN devices as a function of device length ($l$) at several wavelengths (TM: solid curves; TE: dashed curves). Each data point was averaged from the measurements of 3 chips far apart on the wafer. The dotted lines indicate the theoretical maximum responsivities.The responsivity increased by $< 30$\% for devices with $l>$ 50 {\textmu}m.}\label{figS7}
\end{figure}

 Figure \ref{figS7} shows the measured responsivity of PIN PDs vs. $l$ at several wavelengths. The error bars show the maximum and minimum quantities observed across 3 chips, and the data point shows the average.

\section{EQE enhancement for TE at $\lambda $ = 405 nm by narrowing SiN width ($W_{gn}$)}

\begin{figure}[ht]
\centering \includegraphics[width = 0.5 \textwidth]{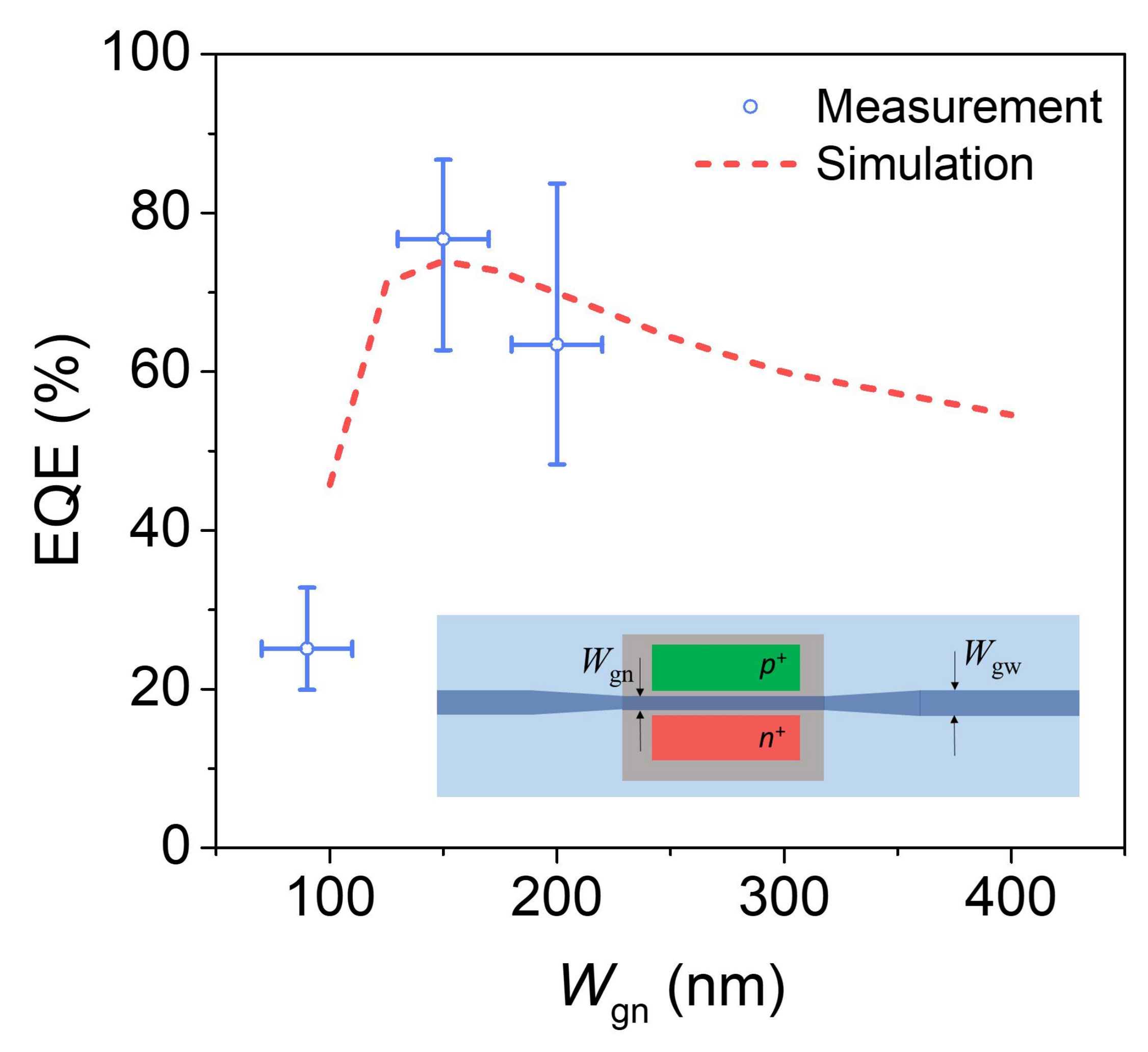}
\caption{Measured and simulated EQE for the SiN-on-Si photodiodes as a function of $W_{gn}$ at $\lambda = 405$ nm for the TE polarized mode. Narrowing the $W_{gn}$ up to 150 nm enhances the evanescent light penetration into Si and consequently the EQE.}\label{fig:S8}
\end{figure}

As seen in Fig. 2(d), the EQE decreased at $\lambda \sim 450$ nm for TE due to the reduced coupling coefficient ($\alpha_{coupling} $) between the SiN waveguide and Si (see Section \ref{sec:absorption}). To enhance $\alpha_{coupling}$, we can narrow $W_{gn}$ to increase the rate of power transfer from  SiN  into Si. Figure \ref{fig:S8} shows this effect of $W_{gn}$ at $\lambda = 405$ nm for the TE polarized mode as an example. As expected, both the measured and simulated EQE exhibited a consistent increase with $W_{gn}$ decreasing down to $\sim$150 nm. A further decrease of $W_{gn}$ resulted in a drastic drop on EQE, likely due to the low mode confinement in SiN causing power scattering into ambient. The simulation indicates a total of  $\sim$19\% EQE enhancement by narrowing $W_{gn}$ from 400 to 150 nm.

\section{Effect of device design dimensions: simulation study}

\begin{figure}[ht]
\centering  \includegraphics[width= 0.75 \textwidth]{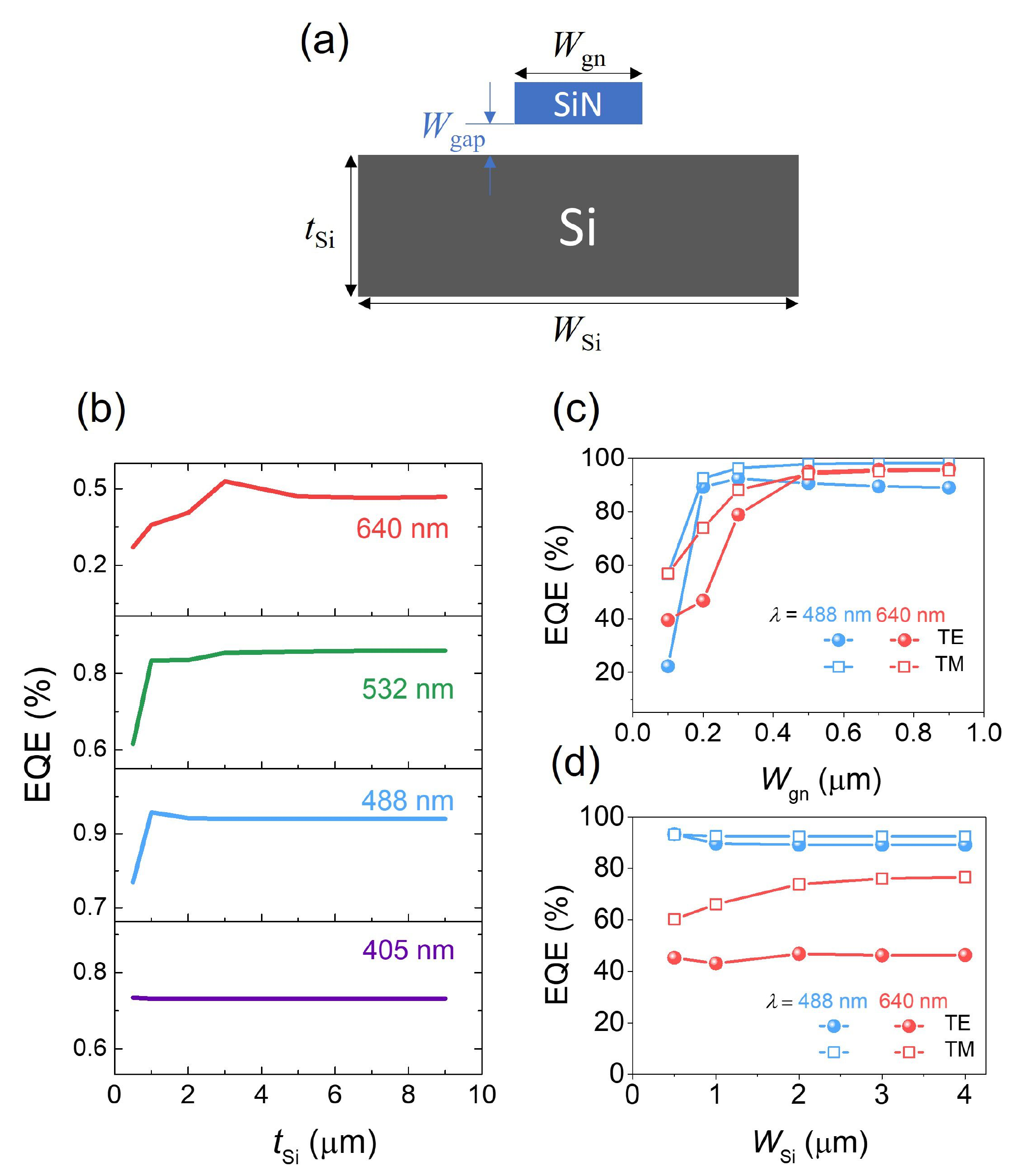} 
\caption{EQE vs. Si thickness, SiN and Si widths. (a) Schematic of SiN-on-Si PD showing the parameters being
studied. (b-d) Simulated EQE as a function of (b) Si thickness ($t_{Si}$) at different visible wavelengths, (c) SiN width ($W_{gn}$)  and (d) Si mesa width ($W_{Si}$) at $\lambda = 488$ and 640 nm.  The device length ($l$) used in the
simulations is 50 {\textmu}m. When held constant, the other geometrical parameters are: $W_{Si} = 2$  {\textmu}m; $W_{gn}  = 200$ nm; $t = 120$ nm; $W_{gap} = 190$ nm; and $t_{Si} = 5$
{\textmu}m for $\lambda \leq 532$ nm and $9$ {\textmu}m for $\lambda > 532$ nm.} \label{fig:S9} 
\end{figure}

\begin{figure}[ht]
\centering  \includegraphics[width=0.8 \textwidth]{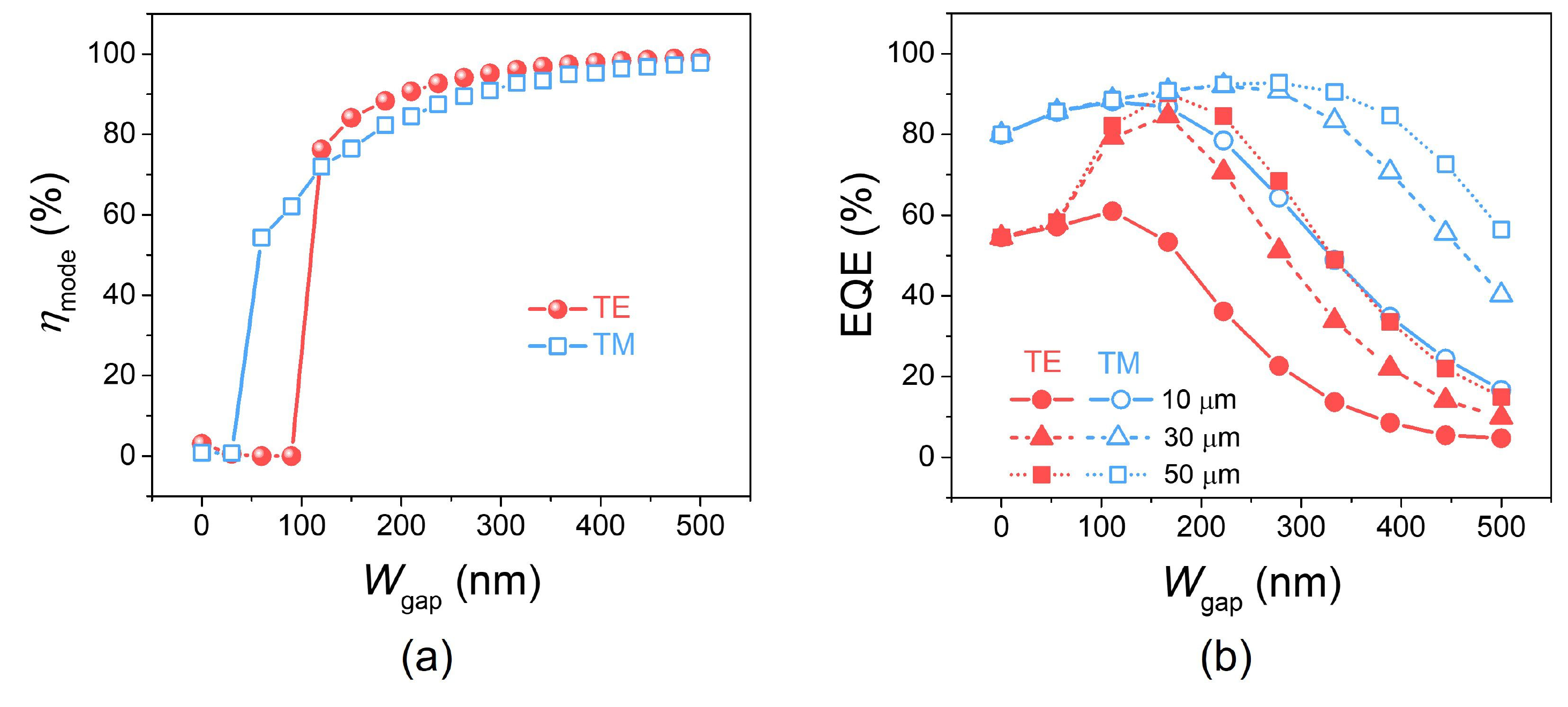} 
\caption{The effect of $W_{gap}$  on (a) $\eta_{mode}$  and (b) EQE at $\lambda = 488 $ nm. The device lengths in (b) are 10, 30 and 50 {\textmu}m, and other design parameters are: $W_{Si}$ = 2 {\textmu}m; $W_{gn} = 200$ nm; $t = 120$ nm; and $t_{Si} = 10$ {\textmu}m. A $W_{gap}$ of $\sim$110-220 nm optimizes the EQE.} \label{fig:S10}
\end{figure}

The EQE was also simulated (see procedures in Section \ref{sec:EQE}) as a function of some key device dimensions. Figure \ref{fig:S9}(a) shows a schematic of the device with these parameters. Figure \ref{fig:S9}(b) shows the effect of Si thickness $t_{Si}$  on EQE. In the simulations, a buried oxide (SiO$_2$) layer of thickness $t_{BOX}$ was added under the Si such that the $t_{Si} + t_{BOX}$ was 5 {\textmu}m for $\lambda \leq 532$ nm and $9$ {\textmu}m for $\lambda > 532$ nm. At $\lambda = 640$ nm, the EQE fluctuates as a function of $t_{Si}$ and flattened as $t_{Si}$ increases. This may be explained by the dependence of the absorption on the phase-matching between the SiN waveguide mode and the slab modes in the thin Si layer. As $t_{Si}$ increases and for shorter wavelengths, more modes within Si are supported, which leads to an EQE that is less sensitive
to $t_{Si}$. At longer wavelengths, Si also has a lower absorption coefficient, which results in a larger penetration depth into Si of the leaky SiN mode. The Si mesa in our PD design (with a height of 2.85 {\textmu}m) results in an EQE that is fairly insensitive to Si thickness.

Figure {\ref{fig:S9}}(c)-(d) show the sensitivity of the EQE to the SiN width ($W_{gn}$) and Si mesa width ($W_{Si}$). The EQE is sensitive to $W_{gn}$ near 200 nm for the two illustrated wavelengths, because the waveguide modes are near or at cut-off (Fig. {\ref{fig:S3}}). Nonetheless, we chose $W_{gn} = 200$ nm to maintain a high EQE at the short wavelengths ($> 60\%$  for $\lambda = 405$ nm). The EQE is insensitive  to $W_{Si} \gtrsim 500$ nm.

Figure {\ref{fig:S10}} shows the effect of $W_{gap}$, the interlayer spacing, on the mode mismatch loss and EQE.  Due to the loss of mode confinement in SiN in the presence of Si, $\eta_{mode} \rightarrow 0$ as $W_{gap} \rightarrow 0$.  $\eta_{mode} \rightarrow 1$ as the Si is separated from the SiN. The EQE peaks at $W_{gap}$ of $\sim 110-220$ nm for both TE and TM polarizations. The EQE decrease is due to the mode mismatch at small $W_{gap}$ and lower evanescent coupling at large $W_{gap}$. To achieve both a high EQE and speed (i.e., at a short device length), a general rule can be to first determine the maximum length that can be used, and then choose a $W_{gap}$ to maximize the EQE.

\section{Junction capacitance and contact resistance extraction for 3-dB OE bandwidth calculation}

The frequency response of PN junction-based photodetectors is determined by both the carrier transit time across the junction and the resistance-capacitance ($RC$) delay of the device \cite{LiuAPL2005}. The carrier transit involves both the drift and diffusion processes and is thus difficult to estimate, while the $RC$ delay estimation is relatively easier, as the device contact resistance $R$ and junction capacitance $C_j$ can be directly extracted from measurements. Calculating the $RC$-limited frequency response helps to understand the dominant factors determining the device 3-dB OE bandwidth. A higher reverse bias ($|V_r|$) leads to both a lower $C_j$ and a shorter carrier transit time, which both contribute to a higher 3-dB OE bandwidth (i.e., shorter FWHM of impulse response seen in Fig. 2(e)). In this section, we show the extracted $R$ and $C_j$ values, estimates of $RC$-limited frequency response, and the comparison with our measurements.

\subsection{Junction capacitance extraction}

\begin{figure}[ht]
\centering   \includegraphics[width = 0.4 \textwidth]{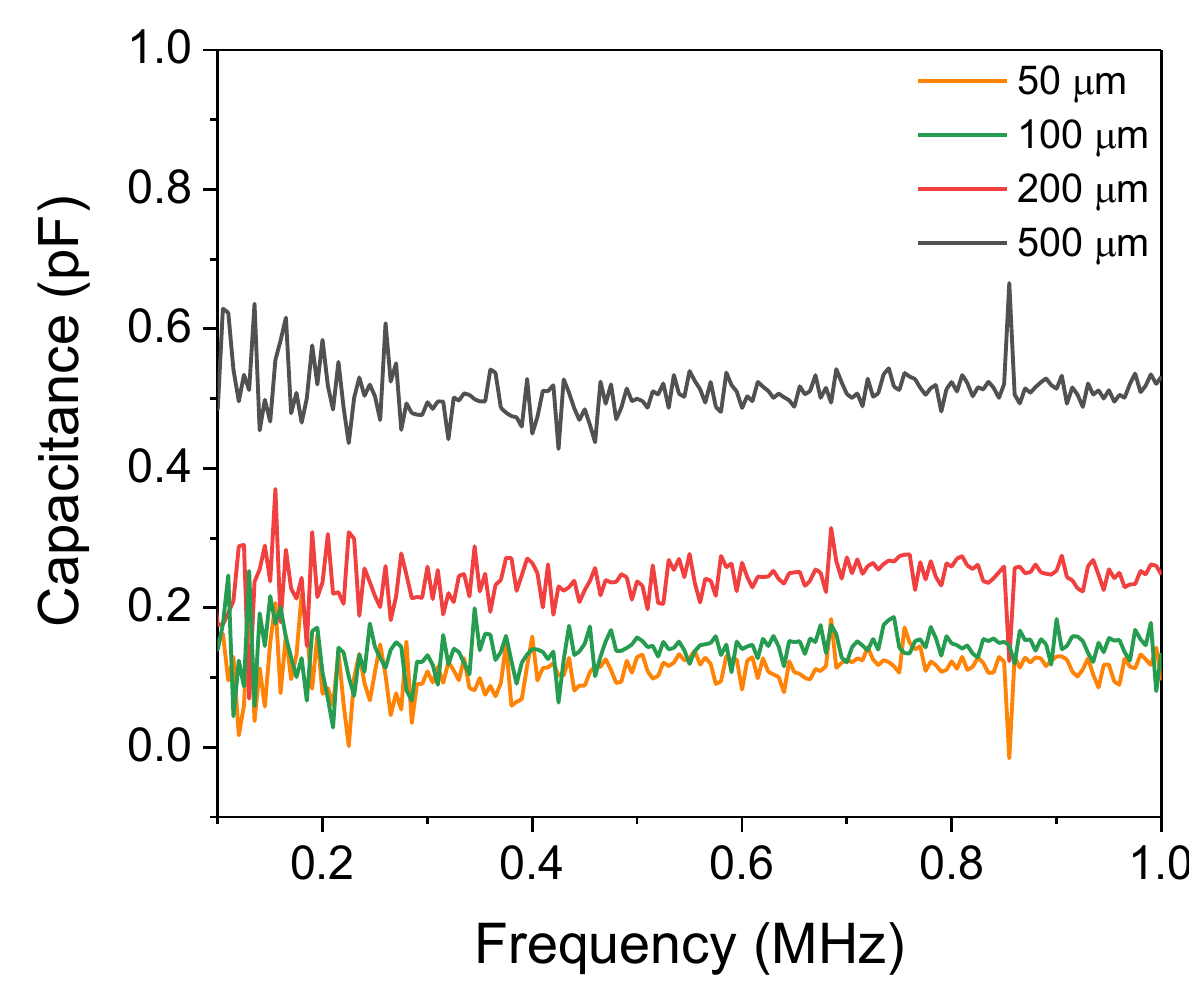} 
\caption{An example plot of measured capacitance ($C$) for PN devices with different lengths at -1 V, as a function of sweeping frequency.}\label{fig:S11}
\end{figure}

\begin{figure}[ht]
\centering  \includegraphics[width= 0.75 \textwidth]{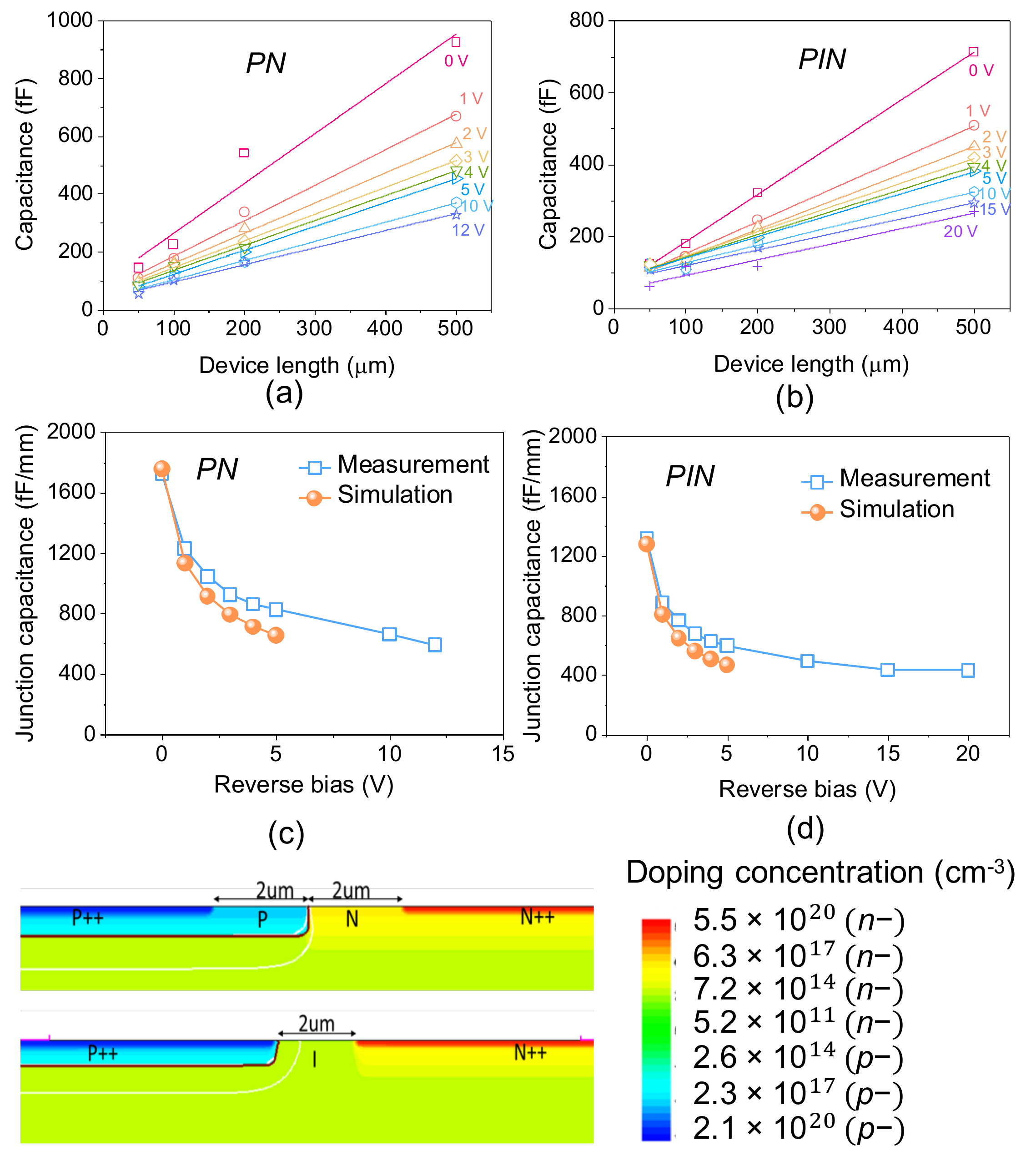} 
\caption{Average measured capacitance ($C$) for (a) PN and (b) PIN  devices as a function of device length at different reverse biases. The corresponding linear fittings resulted in device $C_{sj}$ (slope) and $C_p$ ($y$-intercept). The respective $C_{sj}$  are shown in (c) and (d). (e) Simulated cross-sectional doping profiles for the PN and PIN junctions.}\label{fig:S12}
\end{figure}

Fig. \ref{fig:S11} shows that the capacitance measured by an impedance analyzer is independent of the sweeping frequency at all device lengths. The measurement results have been averaged to reduce noise, and the obtained averages can be linearly fit and extrapolated (see Eq. \ref{eq:S3}) at different reverse biases to extract $C_j$ and parasitic capacitance $C_p$  (Figs. \ref{fig:S12}(a), (b)). $C_j$ scales linearly with device length $l$, while $C_p$  can be
treated as a constant as we applied identical metal wire and contact pad design for all devices. The measured capacitance, $C$ is given by
\begin{equation}\label{eq:S3}
C = C_{sj} \cdot l + C_p,
\end{equation}
where $C_{sj}$  is the junction capacitance per unit length, and $C_j = C_{sj} \cdot l$. Hence, we can extract $C_j$ and $C_p$ from the slopes and $y$-intercepts of the linear fittings and extrapolations, respectively.

From the extrapolation, $C_p = 70 \pm 21$ and $ 55 \pm 17$ fF for PIN and PN  devices, respectively, throughout the measured reverse biases. Figures \ref{fig:S12}(c) and (d) show the extracted $C_{sj}$  as a function of reverse bias for the PN and PIN  device, respectively. The data agrees well with that calculated from the technology computer-aided design (Sentaurus TCAD) simulations of an applied bias of 0 to -5 V, using the implantation and rapid-thermal annealing (RTA) conditions in the fabrication. The Si thickness in the simulation was 4 {\textmu}m to mimic bulk Si. The resulting doping profiles from the TCAD simulations are shown in Fig. \ref{fig:S12}(e).

\subsection{Contact resistance extraction}

\begin{figure}[ht]
\centering  \includegraphics[width=0.75 \textwidth]{ 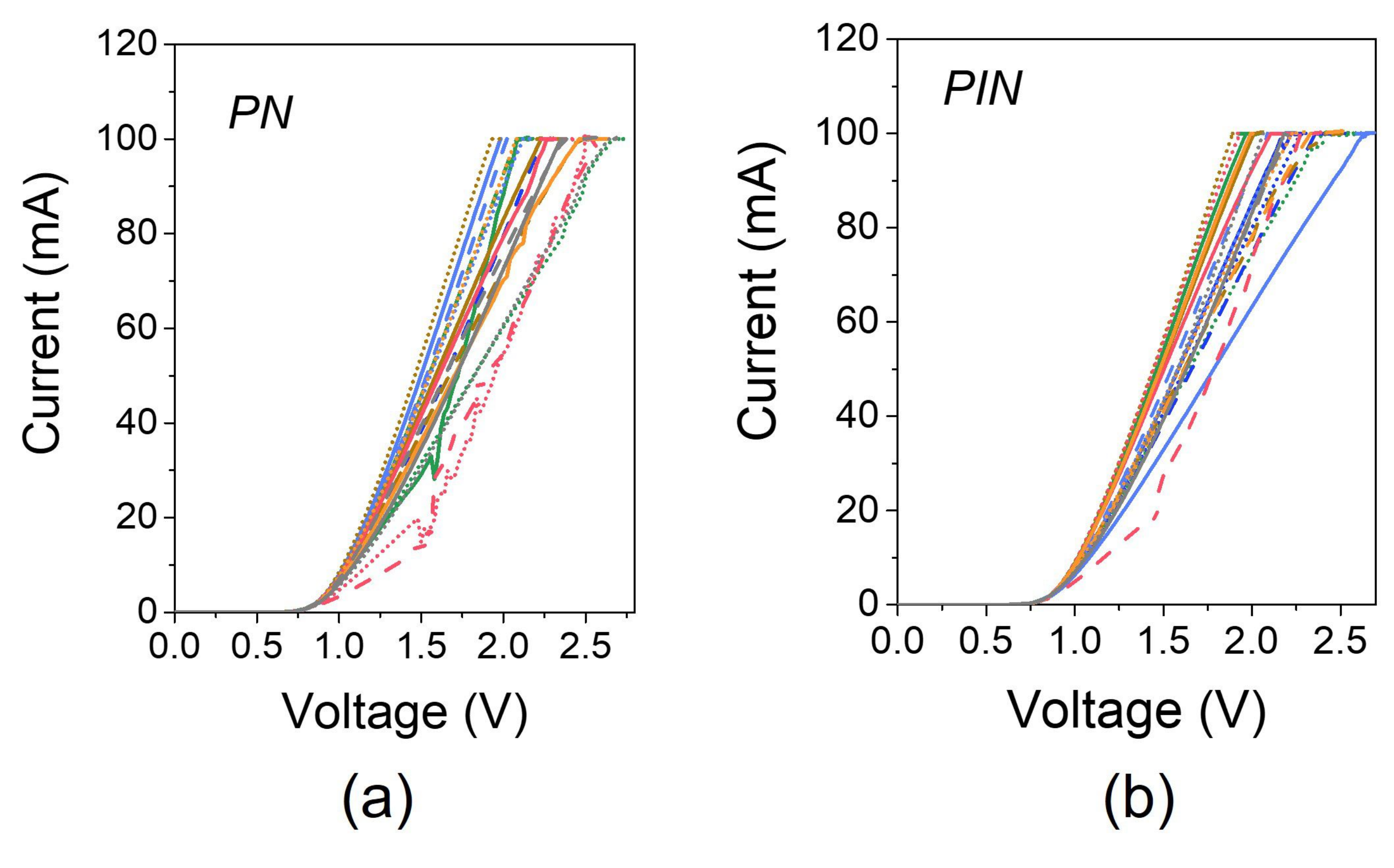} 
\caption{Summary of 50-{\textmu}m long (a) PN and (b) PIN device $I-V$ characteristics (linear scale) at forward bias for the extraction of contact resistance. In total, 21 devices from 7 chips were measured for each device type, which are located uniformly throughout the wafer.} \label{fig:S13}
\end{figure}

Figure \ref{fig:S13} summarizes the $I-V$ characteristics of the 50-{\textmu}m long  PN  and PIN devices in forward bias. Beyond the voltage of 1.5 V, most of the curves exhibit a linear increase up to the compliance current. The contact resistance ($R$) was calculated in this range using $R = \sum_{i=1}^n \frac{dV_i}{dI_i}/n$, were where $n$ is the number of devices tested. We obtained $R = 11.8 \pm 3.9 \mathrm{\Omega} $ and $12.0 \pm 3.0$ $\Omega $ for the PN and PIN devices, respectively. The contact resistance for devices with other lengths were similarly calculated. For the PIN devices, $R = 10.6 \pm 3.5$, $9.0 \pm 3.3$, and $8.7 \pm 3.5$ $\mathrm{\Omega} $ for device lengths of 100, 200
and 500 {\textmu}m, respectively; and for PN devices, the corresponding $R = 11.2 \pm 3.3$,
$8.9 \pm 2.3$, and $9.7 \pm 0.8$ $\mathrm{\Omega }$.

\subsection{$RC$-limited 3-dB bandwidth calculation}

\begin{figure}[ht]
\centering  \includegraphics[width=0.8 \textwidth]{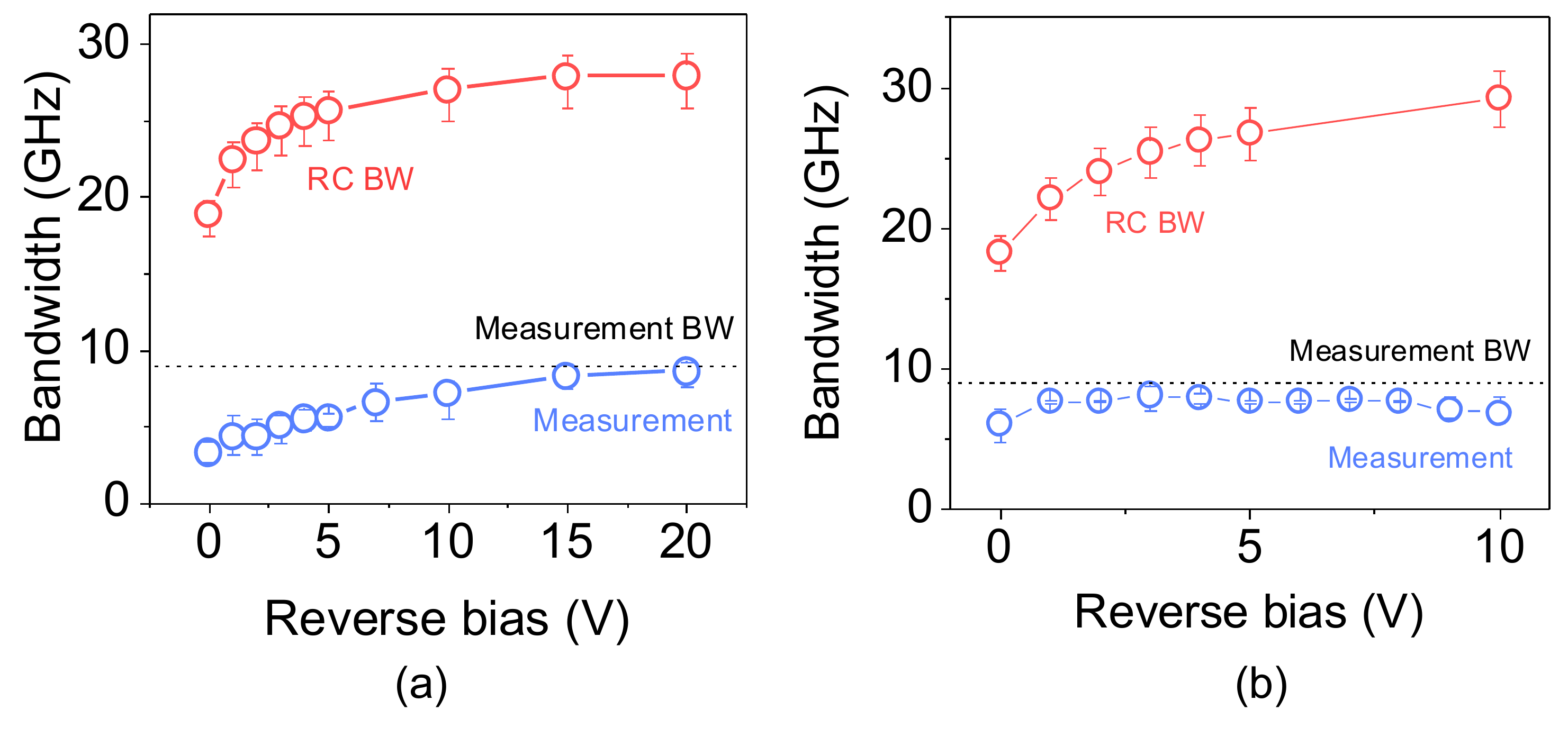} 
\caption{Calculated $RC$-limited bandwidth compared with the measured 3-dB bandwidth for 50-{\textmu}m long (a) PIN and (b) PN devices. Data were collected from 3 different chips on the wafer.}\label{fig:S14}
\end{figure}

Here, we focus on 50-{\textmu}m long devices. The $RC$-limited 3-dB
OE bandwidth can be calculated using \cite{LiuAPL2005, LinPR2017}:
\begin{equation}
f_{RC} = \left[ 2 \pi (R + 50~\Omega) (C_j + C_p) \right]^{-1},
\end{equation}
where $R$, $C_j$, and $C_p$ are the extracted values and 50 $\Omega $ is the load resistance from the measurement
apparatus for impedance matching. Figure \ref{fig:S14} shows the calculated $f_{RC}$  as a function of reverse bias (red curves). For comparison, 3-dB bandwidths obtained from the fast Fourier transform of the impulse responses are included in the same plots (blue curves). The calculated $RC$-limited bandwidths are significantly higher than that from the impulse responses. In the measurements, the bandwidth saturated at $\sim$9 GHz, for both PIN and PN devices, which may be limited by the laser source (pulse FWHM $\sim$50 ps) as well as the oscilloscope bandwidths ($\sim$13 GHz). Thus, for the PIN
devices, at low $V_r$ ($|V_r| < 10$ V), where the 3-dB bandwidth is $< 5$ GHz, the OE bandwidth is likely limited by the carrier transit time; while at high $V_r$ ($|V_r| > 10$ V),  the measurement instrumentation limits the OE bandwidth. For the PN devices, the 3-dB bandwidth is limited by the measurement instrumentations for $|V_r| > 1$ V. A more accurate
measurement of the OE bandwidth would thus require a shorter input pulse width and a higher bandwidth oscilloscope.

\section{APD parameters for $l= 100$ $\mathrm{\mu m}$}

\begin{figure}[ht]
\centering  \includegraphics[width=0.9 \textwidth]{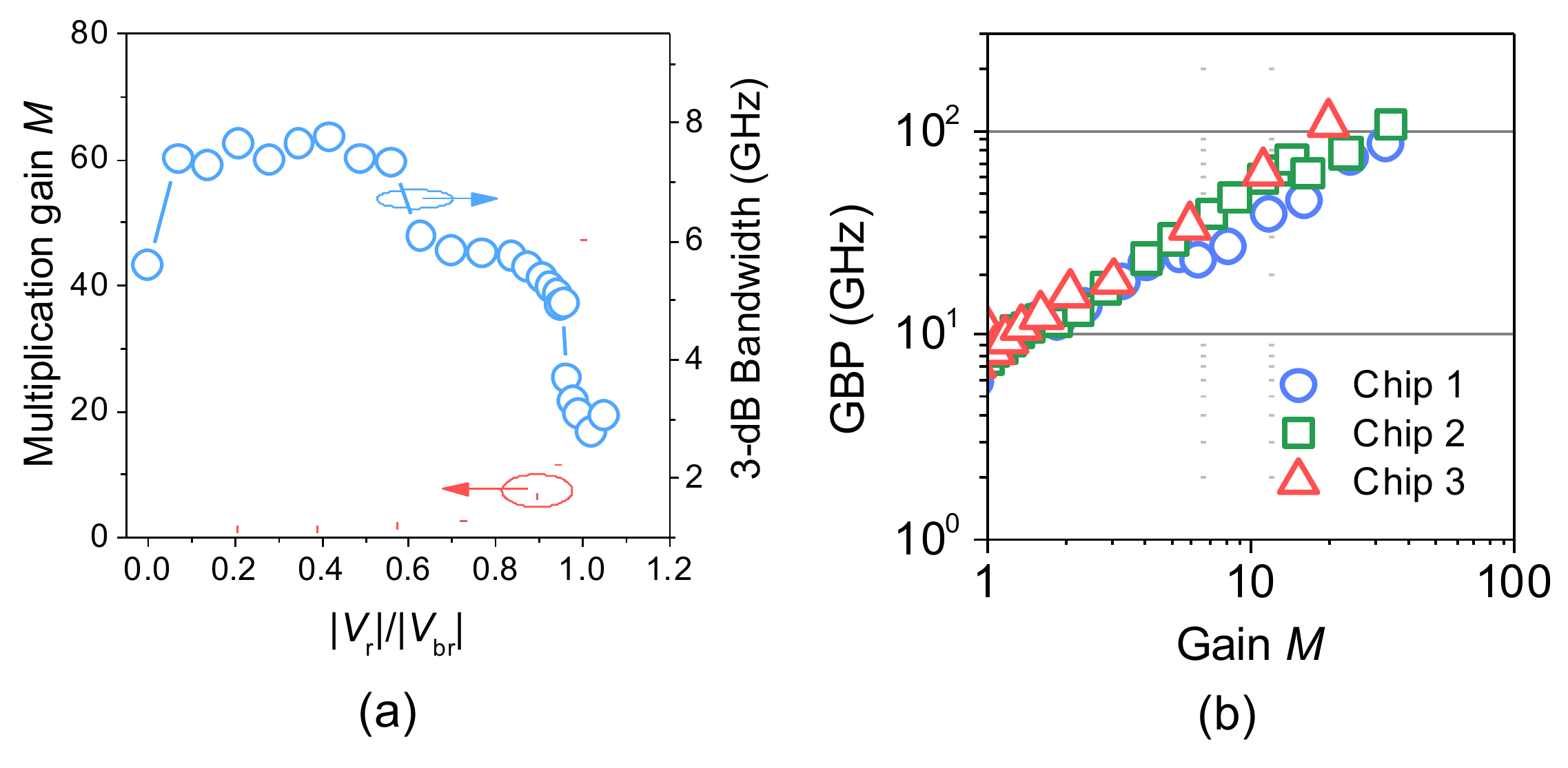} 
\caption{(a) Avalanche multiplication gain ($M$) and 3-dB bandwidth for a PN device with $l = 100$ {\textmu}m. (b) The corresponding GBP as a function of $M$ for devices from 3 different chips on the wafer.}\label{fig:S15}
\end{figure}
Figure \ref{fig:S15} shows  $M$, 3-dB OE bandwidth and GBP results for the 100-{\textmu}m long devices. The method to extract the parameters is described in the Methods section of the main text.

\section{$V_{br}$ determination}

\begin{figure}[ht]
\centering  \includegraphics[width=0.5 \textwidth]{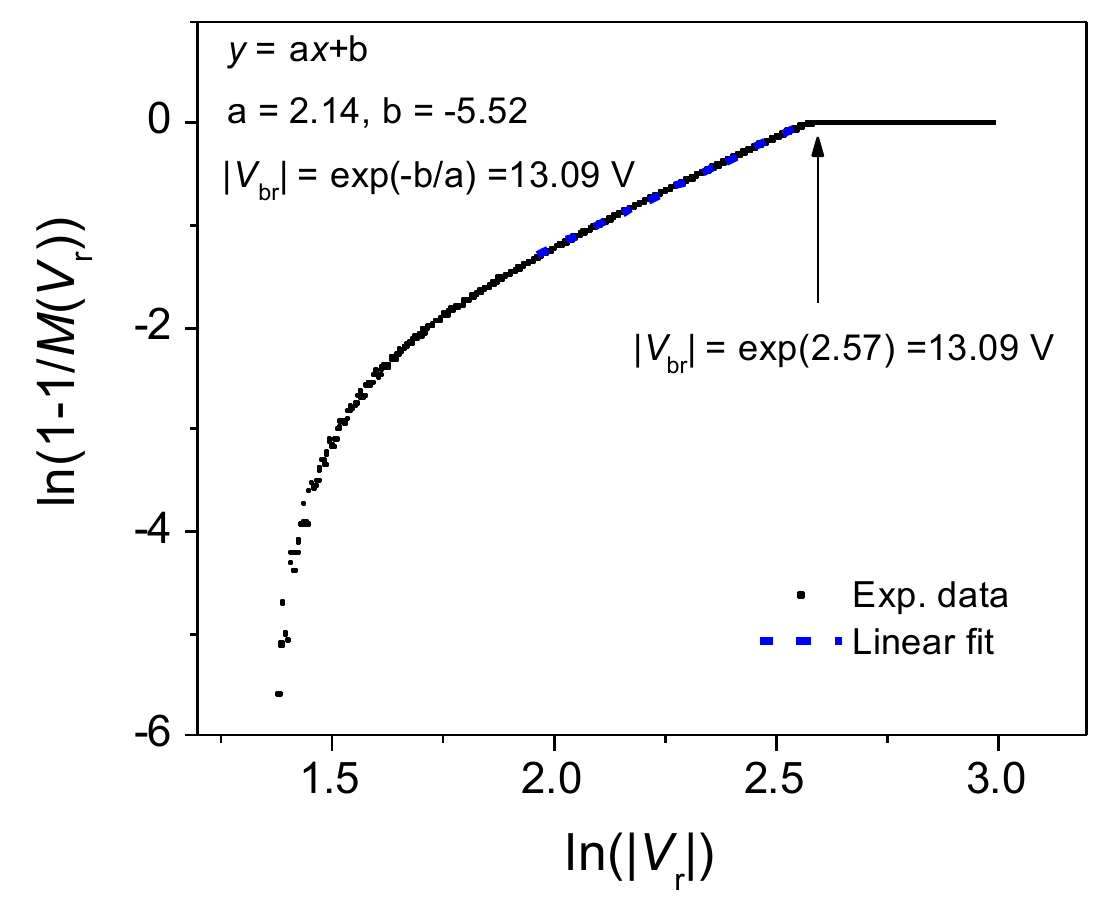} 
\caption{$V_{br}$  extraction via linear fitting and extrapolation.}\label{fig:S16}
\end{figure}

Figure \ref{fig:S16} shows an example plot of the fitting to determine $V_{br}$ for a 50-{\textmu}m long PN device, with a resultant $V_{br} = -13.1$ V. The result agrees well with the onset of the saturation of the curve (indicated by the arrow, $\ln(|V_r|) = 2.57$), beyond which the device clearly experienced an avalanche breakdown (i.e., $\ln \left(1 - \frac{1}{M(V_r)}  \right) \sim 0$).

\section{Comparison with the state of the art}

Table {\ref{table:compare}} compares this work with the state-of-the-art PDs integrated with SiN waveguides operating in the visible and near-infrared (NIR) range. A description of the comparison is in the main manuscript.

\begin{table}[!ht]
\caption{Comparison of SiN waveguide-integrated visible and NIR ($\lambda$ = 400-800 nm) PDs. }\label{table:compare}
\begin{tabular}{ccccccccc}
\hline
\makecell{Device\\  type} & A ($\mu m^{2}$) & $\lambda$ (nm) & $I_{dark}$ (pA) & EQE (\%) & \makecell{BW \\  (GHz)} & $M$ & \makecell{GBP \\  (GHz)} & Ref.\\
\hline
\makecell{SiN-on-Si mesa \\  (PIN, PN)} & 24$\times$50 & 400$\sim$640 & \makecell{144$\pm$42@-5V$^{*}$\\ 266$\pm$65@-15V$^{*}$} & \makecell{$60-88$\\  @-2V$^{*}$} & \makecell{8.6$\pm$1.0\\  @-20V$^{*}$} & $46 \pm 14^{\dagger}$ & $173 \pm 30^{\dagger}$ & \makecell{This\\ work}\\ 
\\
\makecell{End-fire \\  SiN-on-SOI (PN)} & $6^{a}\times 16$ & 685 & $<$70@-2V & $\sim 40^{a}$ & 30 & 12.3 & 234$\pm$25 & \cite{YanikgonulNATCOMM2021s}\\ 
\\
\makecell{SOI-on-SiN \\ (PIN)} & 11.6$\times$200 & 775, 800 & 107@-3V & ~30 & 6 & $\sim 10^{a}$ & 68 & \cite{CuyversOL2022s}\\ 
\\
\makecell{Al$_{2}$O$_{3}$-on-SOI \\  (PIN)} & \makecell{N.A.$\times$100} & 405 & $<$1000 & 76 & - & - & - & \cite{Morgan2021s}\\ 
\\
\makecell{poly-Si\\ (MSM)} & 1.14$\times$10 & 654 & 200@-5V & $67^{a}$ &  - & - & - & \cite{YuanPTL2006}\\ 
\\
\makecell{$\alpha$-Si \\(MSM)} & $30^{a}\times$50 & 660 & \makecell{25@4V,\\  50@8V} &  $0.06^{a}$ & $1 \times 10^{-6}$ & - & - & \cite{DeVitaARXIV2022s}\\ 
\\
MoSe$_{2}$/WS$_{2}$ & $5^{a} \times 13$ & 780 & 50 & \footnotesize{158@-2V}$^{a}$ & 0.02 & - & - & \cite{GherabliARXIV2021s}\\
\\
\makecell{AlGaAs/GaAs\\-on-Ta$_{2}$O$_{5}$} & {\makecell{20$\times$20\\ 20$\times$40}} & 635 & 20@-2V & 22.4 & 12.6 & - & - & \cite{Jafari:22}\\ \hline
\multicolumn{9}{l}{\footnotesize{\textbf{Legend}: $A$: device active area (width$\times$length); $\lambda$: operating wavelength range; $I_{dark}$: dark current;}}\\

\multicolumn{9}{l}{\footnotesize{BW: 3-dB OE bandwidth at unity/low gain. $^{*}$ data from PIN devices; $^{\dagger}$ data from PN devices;}} \\
\multicolumn{9}{l}{\footnotesize{N.A.: not available; $^{a}$ The results were not explicitly reported but inferred from relevant data in literature.}}\\
\end{tabular}
\end{table}

\newpage ~\newpage ~\newpage

%\bibliography{references2}


\end{document}